\documentclass[journal]{IEEEtran}
\usepackage{amsmath,amssymb,amsthm}
\usepackage{booktabs}
\usepackage{graphicx}
\usepackage{algorithm,algorithmic}
\usepackage{cite}
\theoremstyle{definition}   
\newtheorem{lemma}{Lemma}
\newtheorem{theorem}{Theorem}
\newtheorem{corollary}{Corollary}
\newtheorem{proposition}{Proposition}
\newtheorem{remark}{Remark}

\AtBeginDocument{%
  \setlength{\abovedisplayskip}{6pt plus 1pt minus 1pt}%
  \setlength{\belowdisplayskip}{6pt plus 1pt minus 1pt}%
  \setlength{\abovedisplayshortskip}{4pt plus 1pt}%
  \setlength{\belowdisplayshortskip}{4pt plus 1pt}%
}

\begin{document}

\title{Feed-Port-Limited Localization in Pinching-Antenna Cell-Free ISAC}
\author{Changhao~He,~\IEEEmembership{Graduate Student Member,~IEEE}, Xin~He,~\IEEEmembership{Senior Member,~IEEE}, Dawei~Yan, Haoyu Chen,~\IEEEmembership{Senior Member,~IEEE}, and Jiguang~He,~\IEEEmembership{Senior Member,~IEEE}%
\thanks{Changhao He is with the King Abdullah University of Science and Technology (KAUST), Thuwal 23955-6900, Saudi Arabia (e-mail: changhao.he@kaust.edu.sa).}%
\thanks{Xin He is with the School of Computer and Information, Anhui Normal University, Wuhu 241002, China (e-mail: xin.he@ahnu.edu.cn).}%
\thanks{Dawei Yan is with the School of Cyber Security and Computer, Hebei University, Baoding 523106, China (daweiyan@hbu.edu.cn).}%
\thanks{H. Chen is with the Center for Machine Vision and Signal Analysis (CMVS), University of Oulu, FI-90014 Oulu, Finland and Aitomore Oy, Finland (e-mail: chen.haoyu@oulu.fi).}
\thanks{Jiguang He is with the School of Computing and Information Technology, Great Bay University, Dongguan 523000, China, and Great Bay Institute for Advanced Study (GBIAS), Dongguan 523000, China (e-mail: jiguang.he@gbu.edu.cn).~Corresponding author: Jiguang He.}%
}
\maketitle

\begin{abstract}
In a pinching-antenna system (PASS), several radiating elements share a single waveguide feed. We study how this architecture limits device-free target localization in cell-free integrated sensing and communication (ISAC), where passive targets are localized from their echoes as in radar and their scattering coefficients are unknown. We derive an upper bound on the rank of the equivalent Fisher information matrix for the target positions. The bound depends on the transmit covariance rank, the number of receive ports, and the rank of the nuisance derivatives. The bound applies to any number and placement of pinching elements.
Under a signature-rank condition, it identifies when communication-only
transmission yields a singular position information matrix. For a single target with independent feed streams and nonzero link responses, the position information is determined by the energy-weighted variation of the phase and log-amplitude gradients across transmit–receive port pairs. We optimize the transmit covariance for fixed element positions via a tight semidefinite relaxation and update the positions by coordinate search. With matched hardware and feed power under a normalized link budget, the design reduces the worst-target position error bound by 7–25 dB in median paired comparisons with fixed-array, movable-element, and PASS benchmarks. The median penalty for designing without the scattering phases is 0.6 dB. Under the phase-coherent point-scatterer model, distributed ports reduce the bound by 9–19 dB relative to the tested compact co-located deployments.
\end{abstract}

\begin{IEEEkeywords}
Pinching antennas, cell-free ISAC, waveguide feeding, Cram\'er-Rao bound, position error bound, identifiability, nuisance parameters.
\end{IEEEkeywords}

\section{Introduction}\label{sec:intro}

Cell-free integrated sensing and communication (ISAC) networks use
downlink signals from distributed access points (APs) to serve users
and localize passive targets~\cite{liu2022isac}. The localization
considered here is device-free: the targets carry no transceivers and do
not cooperate with the network, and their positions are estimated from
the echoes of the downlink signals, as in radar sensing. The APs observe each
target from different directions, providing the geometric diversity
needed for multistatic localization~\cite{behdad2024multistatic,wang2026cfisac}.
Reconfiguring the aperture at each AP can further improve sensing, but
an array with a dedicated radio-frequency (RF) chain per element becomes
more costly as the number of elements grows.

In a pinching-antenna system (PASS), several movable radiating elements
share a single dielectric waveguide feed~\cite{fukuda2022pinching,ding2025flexible,liu2026survey}.
Adding or moving elements changes how the waveguide illuminates the
scene. All elements on a transmit waveguide still share the same feed
signal, and each receive segment delivers a single observation through
its output port. When the target scattering coefficients are unknown,
some changes in the received echo can be explained by changes in those
coefficients. We study how these two features constrain localization
and how the placement of ports and elements affects the position
information that remains.

Following the hardware demonstration in~\cite{fukuda2022pinching},
subsequent work analyzed the communication performance of
PASS~\cite{ding2025flexible,liu2026survey} and showed how waveguide
attenuation affects element placement~\cite{xu2026attenuation}.
Segmented and center-fed architectures expand the available
communication degrees of freedom by providing additional feed
ports~\cite{ouyang2026swan,gan2026cpass}.

Prior studies have characterized range estimation bounds and designed
PASS for sensing and detection~\cite{ding2025crlb,wang2025sensing,mao2026multi}.
Other work has optimized beamforming, element placement, target
scheduling, or power allocation using estimation
bounds~\cite{li2026pass_isac,jiang2026bcrb,geng2026swan,pakravan2026dual}.
With an unknown reflection coefficient, the single-target position error bound for one element per waveguide has been factored into transmit illumination and receive geometry in \cite{bozanis2026secure}. A projected Jacobian has been used in \cite{yang2026joint} to characterize identifiability and jointly design the element positions and transmit covariance in a near-field multicast system with a co-located receive array.

Fluid and movable antenna systems exploit element placement to
improve sensing~\cite{wong2021fluid,zhu2024movable,ma2024ma_sensing}.
In cell-free ISAC, distributed observations provide additional
geometric diversity for reconfigurable apertures~\cite{he2026crb}.
This principle is well established in distributed multiple-input
multiple-output radar~\cite{godrich2010target}. The equivalent Fisher
information matrix (EFIM) provides a way to quantify the position
information after accounting for nuisance parameters~\cite{shen2010fundamental}.

The existing PASS analyses explain how a single-site aperture shapes
sensing information. In a cell-free deployment, independent transmit
streams and distributed receive ports determine the available
observation dimensions, while the unknown scattering coefficients
consume part of those dimensions. In this paper, we determine when
communication streams provide enough observation dimensions for
multi-target localization, and how distributing the ports changes the
position information when the links share an unknown scattering
coefficient.

The main contributions of this paper are as follows.
\begin{itemize}
\item
We bound the rank of the position EFIM using the number of independent
transmit streams, the number of receive ports, and the dimension of the
nuisance subspace. The bound applies to any number and placement of
pinching elements. For fixed hardware dimensions and transmit rank, a
single configuration that attains the bound with a nonsingular nuisance
block establishes generic tightness. With calibrated waveguides and
maximal-rank target signatures, communication-only transmission yields
a singular position EFIM whenever the product of the user count and
the receive-port count is less than twice the target count.

\item
For a single target, we separate the position EFIM into per-link
contributions and a cooperative contribution from the shared unknown
scattering coefficient. With independent feed streams and nonzero link
responses, the per-link contributions vanish. The position information
then comes from the energy-weighted variation of phase and log-amplitude
gradients across transmit--receive port pairs.

\item
Appending elements to a fixed layout changes the effective channel and
its position derivatives by a geometrically decaying amount. Under
uniform conditioning of the nuisance Jacobian and the position EFIM,
the same decay holds for changes in the position error bound (PEB).
We also characterize how waveguide attenuation affects the position
EFIM.

\item
We minimize the worst-target PEB subject to per-user
signal-to-interference-plus-noise ratio (SINR) requirements and per-AP
feed-power limits. For fixed element positions, a tight semidefinite
relaxation solves the covariance subproblem with lossless rank-one
recovery of the user covariances. We update the element positions by
monotone coordinate search with steps scaled to the guided wavelength.
\end{itemize}

\emph{Notations:} Bold lower-case and upper-case letters denote vectors and matrices. $(\cdot)^\top$, $(\cdot)^*$ and $(\cdot)^H$ denote transpose, conjugate and conjugate transpose, $\otimes$ the Kronecker product, $\operatorname{vec}(\cdot)$ column stacking, $j=\sqrt{-1}$, $\operatorname{Re}\{\cdot\}$ and $\operatorname{Im}\{\cdot\}$ the real and imaginary parts, $\operatorname{tr}(\cdot)$ the trace, $\mathbb{E}[\cdot]$ the expectation, $\|\cdot\|$ the Euclidean norm, $(\cdot)^\dagger$ the Moore--Penrose pseudoinverse, $\mathbf{I}$ the identity matrix, $\operatorname{diag}(\cdot)$ and $\mathrm{blkdiag}(\cdot)$ (block) diagonal matrices, $\lambda_{\min}(\cdot)$ and $\lambda_{\max}(\cdot)$ the smallest and largest eigenvalues, $\mathcal{CN}(0,\sigma^2)$ the circularly symmetric complex Gaussian distribution with variance $\sigma^2$, $\mathbf{P}^\perp_{\mathbf{A}}$ the orthogonal projector onto the complement of the column space of $\mathbf{A}$, and $\mathbf{A}\succeq\mathbf{0}$ a positive semidefinite matrix.

Sections~\ref{sec:model} and~\ref{sec:problem} describe the system model and formulate the joint design problem, respectively. Section~\ref{sec:limits} derives the feed-port limits on target localization, and Section~\ref{sec:algorithm} presents the two-timescale design algorithm. Section~\ref{sec:numerical} reports the numerical results, and Section~\ref{sec:conclusion} concludes the paper.

\section{Waveguide-Fed Cell-Free ISAC Model}\label{sec:model}
This section models the downlink signals and target echoes in a waveguide-fed cell-free network. It defines the position error bound for three cases that differ in which scattering and waveguide parameters are known.

\begin{figure*}[t]
\centering
\includegraphics[width=\textwidth]{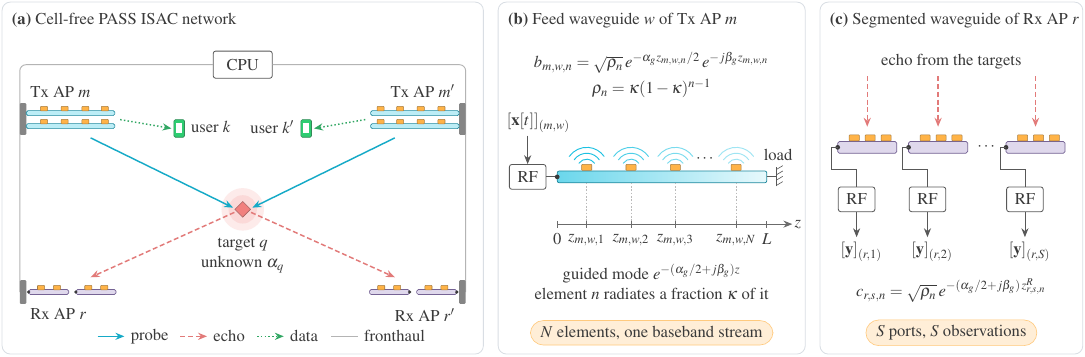}
\vspace{-15pt}
\caption{System model. (a) Tx APs with $W$ feed waveguides illuminate targets and serve users; Rx APs with segmented waveguides collect the echoes. (b) One Tx waveguide: a single RF injects one baseband stream, element $n$ radiates with coefficient $b_{m,w,n}$ in \eqref{eq:b}, and the fraction $(1-\kappa)^Ne^{-\alpha_gL}$ of the feed power reaches the matched load. (c) One Rx AP: each of the $S$ segments has its own output port.}
\label{fig:system}
\end{figure*}

\subsection{Network Geometry}\label{sec:geometry}

Consider a cell-free ISAC network with $M_T$ transmit access points (Tx APs), $M_R$ receive APs (Rx APs), $K$ single-antenna downlink users, and $Q$ passive point targets, all coordinated by a central processing unit (CPU) as illustrated in Fig.~\ref{fig:system}(a). All APs operate at carrier wavelength $\lambda$ and are mounted at a common height $H$, while the users and targets lie on the ground plane. Target $q$ is located at $\mathbf{p}_q=[x_q,y_q]^\top\in\mathbb{R}^2$ with three-dimensional (3D) coordinates $\tilde{\mathbf{p}}_q=[\mathbf{p}_q^\top,0]^\top$, and user $k$ is located at $\mathbf{u}_k\in\mathbb{R}^2$ with 3D coordinates $\tilde{\mathbf{u}}_k=[\mathbf{u}_k^\top,0]^\top$.

\subsubsection{Tx AP}
Each Tx AP $m\in\mathcal{M}_T=\{1,\dots,M_T\}$ is equipped with $W$ dielectric waveguides. Waveguide $w\in\mathcal{W}=\{1,\dots,W\}$ is a straight segment of length $L$ extending from the feed point $\mathbf{o}_{m,w}\in\mathbb{R}^3$ along the horizontal unit vector $\mathbf{d}_{m,w}$. Fed at one end by a dedicated RF chain, each waveguide carries $N$ pinching elements. The $n$-th element is placed at waveguide coordinate $z_{m,w,n}\in[0,L]$, corresponding to the spatial position
\begin{equation}
\boldsymbol{\psi}_{m,w,n}(z_{m,w,n})=\mathbf{o}_{m,w}+z_{m,w,n}\,\mathbf{d}_{m,w}.
\label{eq:pinch_pos}
\end{equation}
Elements along each waveguide are ordered with a minimum spacing $d_{\min}$:
\begin{equation}
0\le z_{m,w,1},\; z_{m,w,n+1}-z_{m,w,n}\ge d_{\min},\; z_{m,w,N}\le L .
\label{eq:order}
\end{equation}
Stacking the coordinates gives $\mathbf{z}_{m,w}=[z_{m,w,1},\dots,z_{m,w,N}]^\top$, $\mathbf{z}_m=[\mathbf{z}_{m,1}^\top,\dots,\mathbf{z}_{m,W}^\top]^\top$, and $\mathbf{z}_T=[\mathbf{z}_1^\top,\dots,\mathbf{z}_{M_T}^\top]^\top\in\mathbb{R}^{M_TWN}$, whose feasible set defined by \eqref{eq:order} across all $(m,w)$ is denoted by $\mathcal{Z}_T$.

\subsubsection{Rx AP}
Each Rx AP $r\in\mathcal{M}_R=\{1,\dots,M_R\}$ employs a segmented waveguide comprising $S$ electrically isolated segments, which prevents re-radiation along a long receive guide and keeps the output ports independent. Segment $s\in\mathcal{S}=\{1,\dots,S\}$ has a dedicated output port (one receive RF chain), length $L_R$, origin $\mathbf{o}^R_{r,s}$, direction $\mathbf{d}^R_{r,s}$, and $N_R$ pinching elements at coordinates $\mathbf{z}^R_{r,s}\in\mathbb{R}^{N_R}$ subject to the ordering constraint in \eqref{eq:order} (with $L_R$ and $N_R$ in place of $L$ and $N$). The spatial positions $\boldsymbol{\psi}^R_{r,s,n}$ follow \eqref{eq:pinch_pos}. Collecting all receive coordinates yields $\mathbf{z}_R\in\mathbb{R}^{M_RSN_R}$ with feasible set $\mathcal{Z}_R$, and we denote the joint coordinate vector by $\mathbf{z}=[\mathbf{z}_T^\top,\mathbf{z}_R^\top]^\top\in\mathcal{Z}=\mathcal{Z}_T\times\mathcal{Z}_R$.

\subsection{Waveguide Feeding Model}\label{sec:feed}

A baseband signal injected at the feed of waveguide $(m,w)$ propagates as a guided mode with phase constant $\beta_g=2\pi n_{\mathrm{eff}}/\lambda$ and power attenuation constant $\alpha_g$ (Np/m), where $n_{\mathrm{eff}}$ is the effective refractive index. Each pinching element couples a fraction $\kappa\in(0,1)$ of the remaining guided power into free space. Since the elements are ordered along the waveguide, the guided power reaching element $n$ has already been depleted by elements $1,\dots,n-1$, yielding the complex radiation coefficient illustrated in Fig.~\ref{fig:system}(b)
\begin{equation}
b_{m,w,n}(z_{m,w,n})=\sqrt{\rho_n}\;e^{-\alpha_g z_{m,w,n}/2}\;e^{-j\beta_g z_{m,w,n}},
\label{eq:b}
\end{equation}
with $\rho_n=\kappa(1-\kappa)^{n-1}$. The coupling fraction $\kappa$ is taken to be independent of the element position, and mutual coupling between radiating elements and re-radiation into the guide are neglected. Collecting $\mathbf{b}_{m,w}(\mathbf{z}_{m,w})=[b_{m,w,1},\dots,b_{m,w,N}]^\top\in\mathbb{C}^N$, we define the AP feeding matrix $\mathbf{B}_m(\mathbf{z}_m)=\mathrm{blkdiag}(\mathbf{b}_{m,1},\dots,\mathbf{b}_{m,W})\in\mathbb{C}^{WN\times W}$. The waveguide radiation efficiency, defined as the fraction of feed power radiated into free space, is
\begin{equation}
\eta_{m,w}(\mathbf{z}_{m,w})=\|\mathbf{b}_{m,w}\|^2=\sum_{n=1}^{N}\rho_n e^{-\alpha_g z_{m,w,n}}<1 .
\label{eq:eta}
\end{equation}
A fraction $(1-\kappa)^Ne^{-\alpha_gL}$ of the feed power reaches the matched termination, while the remainder is dissipated along the guide. By reciprocity, each receive segment follows the same model with the roles of feed and termination reversed: the signal coupled at element $n$ of segment $(r,s)$ propagates a distance $z^R_{r,s,n}$ to the output port, resulting in the receive coefficient and AP combining matrix
\begin{align}
c_{r,s,n}(z^R_{r,s,n})&=\sqrt{\rho_n}\;e^{-\alpha_g z^R_{r,s,n}/2}\;e^{-j\beta_g z^R_{r,s,n}},\notag\\
\mathbf{C}_r&=\mathrm{blkdiag}\big(\mathbf{c}_{r,1},\dots,\mathbf{c}_{r,S}\big)\in\mathbb{C}^{SN_R\times S},
\label{eq:Cr}
\end{align}
where $\mathbf{c}_{r,s}=[c_{r,s,1},\dots,c_{r,s,N_R}]^\top\in\mathbb{C}^{N_R}$.

\begin{remark}[Feed-port limitation]\label{rem:feed}
By construction, $\operatorname{rank}\mathbf{B}_m=W$ for all $\mathbf{z}_m\in\mathcal{Z}_T$, regardless of $N$. The $WN$ radiating elements of a Tx AP are driven by $W$ feed ports, supporting at most $W$ independent baseband streams, while the $SN_R$ elements of an Rx AP deliver $S$ scalar observations. The number of independent streams actually excited is determined by the transmit waveform through $\operatorname{rank}(\mathbf{R}_x)$ in \eqref{eq:Rx}. Adjusting $\mathbf{z}$ reshapes the columns of $\mathbf{B}_m$ and $\mathbf{C}_r$ without increasing their rank, in contrast to fluid or movable arrays that dedicate one RF chain to each element.
\end{remark}

\subsection{Transmit Signal}\label{sec:tx}

The targets are observed over a coherent processing interval (CPI) of $T$ symbols, which is the sensing counterpart of a coherence block in block fading: the target positions and scattering coefficients stay constant within one CPI, with no Doppler shift, so that the echoes of the $T$ symbols can be processed coherently. The CPI is assumed to be no longer than the channel coherence time, so that the user channels are also constant within it. In each CPI, Tx AP $m$ injects into its $W$ waveguides the baseband vector
\begin{equation}
\mathbf{x}_m[t]=\sum_{k=1}^{K}\mathbf{v}_{m,k}\,s_k[t]+\tilde{\mathbf{s}}_m[t]\in\mathbb{C}^W,
\;\; t=1,\dots,T,
\label{eq:xm}
\end{equation}
where $s_k[t]$ is the unit-power data symbol of user $k$, $\mathbf{v}_{m,k}\in\mathbb{C}^W$ is the corresponding precoder at AP $m$, and $\tilde{\mathbf{s}}_m[t]$ is a dedicated sensing waveform with $\mathbb{E}[\tilde{\mathbf{s}}_m\tilde{\mathbf{s}}_{m'}^H]=\mathbf{S}_{mm'}$. Stacking over APs, $\mathbf{x}[t]=[\mathbf{x}_1^\top[t],\dots,\mathbf{x}_{M_T}^\top[t]]^\top\in\mathbb{C}^{M_TW}$, $\mathbf{v}_k=[\mathbf{v}_{1,k}^\top,\dots,\mathbf{v}_{M_T,k}^\top]^\top$, $\mathbf{Q}_k=\mathbf{v}_k\mathbf{v}_k^H$, $\mathbf{S}=[\mathbf{S}_{mm'}]\succeq\mathbf{0}$, and the transmit covariance is
\begin{equation}
\mathbf{R}_x=\frac{1}{T}\sum_{t=1}^{T}\mathbf{x}[t]\mathbf{x}^H[t]\;\approx\;\sum_{k=1}^{K}\mathbf{Q}_k+\mathbf{S}\in\mathbb{C}^{M_TW\times M_TW}.
\label{eq:Rx}
\end{equation}
The large-$T$ approximation in \eqref{eq:Rx} is treated as exact in the design. Let $\mathbf{E}_m\in\{0,1\}^{M_TW\times M_TW}$ select the $W$ waveguides of AP $m$. Amplifiers sit at the feed, so the per-AP budget applies to the feed power, $\operatorname{tr}(\mathbf{E}_m\mathbf{R}_x)\le P_m$.

\subsection{Sensing Observation}\label{sec:sensing}

Free-space propagation between a pinching element at $\boldsymbol{\psi}$ and a ground point $\tilde{\mathbf{p}}$ is modeled by the spherical-wave response
\begin{equation}
g(\tilde{\mathbf{p}},\boldsymbol{\psi})=\frac{\sqrt{\eta_0}}{\|\tilde{\mathbf{p}}-\boldsymbol{\psi}\|}\,
\exp\!\Big(-j\frac{2\pi}{\lambda}\|\tilde{\mathbf{p}}-\boldsymbol{\psi}\|\Big),
\label{eq:g}
\end{equation}
where $\eta_0$ accounts for the element gain and the free-space path-loss constant. The signal bandwidth is assumed to be small enough that \eqref{eq:g} holds across the band. We define the Tx element-to-target vector $\mathbf{g}_{m,q}\in\mathbb{C}^{WN}$ with entries $[\mathbf{g}_{m,q}]_{(w,n)}=g(\tilde{\mathbf{p}}_q,\boldsymbol{\psi}_{m,w,n})$ and the Rx target-to-element vector $\mathbf{g}^R_{r,q}\in\mathbb{C}^{SN_R}$ with entries $[\mathbf{g}^R_{r,q}]_{(s,n)}=g(\tilde{\mathbf{p}}_q,\boldsymbol{\psi}^R_{r,s,n})$. The effective Tx and Rx channels seen at the waveguide ports are
\begin{equation}
\mathbf{h}^T_{m,q}=\mathbf{B}_m^\top\mathbf{g}_{m,q}\in\mathbb{C}^{W},
\;\;
\mathbf{h}^R_{r,q}=\mathbf{C}_r^\top\mathbf{g}^R_{r,q}\in\mathbb{C}^{S}.
\label{eq:heff}
\end{equation}
Stacking these vectors across APs yields $\mathbf{h}^T_q=[\mathbf{h}^{T\top}_{1,q},\dots,\mathbf{h}^{T\top}_{M_T,q}]^\top\in\mathbb{C}^{M_TW}$ 
and $\mathbf{h}^R_q=[\mathbf{h}^{R\top}_{1,q},\dots,\mathbf{h}^{R\top}_{M_R,q}]^\top\in\mathbb{C}^{M_RS}$. Target $q$ is characterized by an unknown complex scattering coefficient $\alpha_q$ that incorporates its radar cross section and bistatic phase. Each target is modeled as an isotropic point scatterer whose coefficient $\alpha_q$ is common to all Tx--Rx AP pairs, as in coherent distributed MIMO radar~\cite{godrich2010target}. All APs are phase-synchronized at the CPU, so the echoes of different links are coherent, and direct Tx-to-Rx leakage and clutter are assumed to be suppressed before the echoes are processed. For an extended target whose reflectivity changes with aspect angle, each Tx--Rx pair has its own coefficient and the information gained from the shared coefficient (Section~\ref{sec:coop}) vanishes, meaning the reported PEB is optimistic for such targets. Over the CPI, the echoes collected across all Rx segments form the matrix
\begin{equation}
\mathbf{Y}=\sum_{q=1}^{Q}\alpha_q\,\mathbf{h}^R_q\,\mathbf{h}^{T\top}_q\,\mathbf{X}+\mathbf{N}\in\mathbb{C}^{M_RS\times T},
\label{eq:Y}
\end{equation}
where $\mathbf{X}=[\mathbf{x}[1],\dots,\mathbf{x}[T]]$ is known at the CPU and $\mathbf{N}$ has i.i.d.\ $\mathcal{CN}(0,\sigma^2)$ entries, since the Rx segments are electrically isolated. Vectorizing $\mathbf{Y}$ gives
\begin{equation}
\mathbf{y}=\operatorname{vec}(\mathbf{Y})=\sum_{q=1}^{Q}\alpha_q\,\mathbf{s}_q(\mathbf{p}_q,\mathbf{z})+\mathbf{n},
\;
\mathbf{s}_q=\big(\mathbf{X}^\top\mathbf{h}^T_q\big)\otimes\mathbf{h}^R_q ,
\label{eq:y}
\end{equation}
where $\mathbf{s}_q\in\mathbb{C}^{M_RST}$ is the space-time signature of target $q$ and $\mathbf{n}=\operatorname{vec}(\mathbf{N})$.

\subsection{Communication Observation}\label{sec:comm}

User $k$ sees the downlink channel $\mathbf{h}_k=[\mathbf{h}_{1,k}^\top,\dots,\mathbf{h}_{M_T,k}^\top]^\top\in\mathbb{C}^{M_TW}$ with $\mathbf{h}_{m,k}=\mathbf{B}_m^\top\mathbf{g}_{m,k}$ and $[\mathbf{g}_{m,k}]_{(w,n)}=g(\tilde{\mathbf{u}}_k,\boldsymbol{\psi}_{m,w,n})$. Its received symbol is $y_k[t]=\mathbf{h}_k^\top\mathbf{x}[t]+n_k[t]$, $n_k\sim\mathcal{CN}(0,\sigma_k^2)$, and its signal-to-interference-plus-noise ratio (SINR) is
\begin{equation}
\mathrm{SINR}_k(\mathbf{z}_T,\{\mathbf{Q}_i\},\mathbf{S})=
\frac{\mathbf{h}_k^\top\mathbf{Q}_k\mathbf{h}_k^*}
{\sum_{i\neq k}\mathbf{h}_k^\top\mathbf{Q}_i\mathbf{h}_k^*+\mathbf{h}_k^\top\mathbf{S}\mathbf{h}_k^*+\sigma_k^2}.
\label{eq:sinr}
\end{equation}
The sensing waveform is treated as interference at the users. Perfect downlink channel state information (CSI) $\{\mathbf{h}_k\}$ is assumed to be available at the CPU for the commanded coordinates $\mathbf{z}$.

\subsection{Unknown Parameters and Three Nested Models}\label{sec:nuisance}

At the estimation stage, the target positions $\mathbf{p}=[\mathbf{p}_1^\top,\dots,\mathbf{p}_Q^\top]^\top\in\mathbb{R}^{2Q}$ are the parameters of interest, while the complex scattering coefficients $\boldsymbol{\alpha}=[\alpha_1,\dots,\alpha_Q]^\top\in\mathbb{C}^Q$ and, depending on the calibration state, the waveguide constants $\boldsymbol{\theta}_g=[\alpha_g,\beta_g]^\top\in\mathbb{R}^2$ act as nuisance parameters. Denoting the complete vector of real unknowns by $\boldsymbol{\xi}$, we consider three nested parameterizations.
\emph{M-I (Ideal):} $\boldsymbol{\xi}=\mathbf{p}$, with both $\boldsymbol{\alpha}$ and $\boldsymbol{\theta}_g$ assumed known.
\emph{M-II (Nuisance-aware):} $\boldsymbol{\xi}=[\mathbf{p}^\top,\operatorname{Re}\boldsymbol{\alpha}^\top,\operatorname{Im}\boldsymbol{\alpha}^\top]^\top$, with $\boldsymbol{\alpha}$ unknown and $\boldsymbol{\theta}_g$ known.
\emph{M-III (Calibration-aware):} $\boldsymbol{\xi}=[\mathbf{p}^\top,\operatorname{Re}\boldsymbol{\alpha}^\top,\operatorname{Im}\boldsymbol{\alpha}^\top,\boldsymbol{\theta}_g^\top]^\top$, with both $\boldsymbol{\alpha}$ and $\boldsymbol{\theta}_g$ unknown, where the constants $\boldsymbol{\theta}_g$ are common to all waveguides of the network.

When element placement errors are present, the actual coordinates $\mathbf{z}^{\mathrm{act}}=\mathbf{z}+\boldsymbol{\delta}_z$ remain fixed over the CPI, where $\boldsymbol{\delta}_z$ follows a known distribution. Rather than being estimated as part of $\boldsymbol{\xi}$, $\boldsymbol{\delta}_z$ is treated as a random deployment perturbation on which the performance bound is conditioned.

\subsection{Performance Metric}\label{sec:metric}

Let $\boldsymbol{\mu}(\boldsymbol{\xi})=\sum_{q=1}^{Q}\alpha_q\mathbf{s}_q$ denote the noise-free mean of $\mathbf{y}$ in \eqref{eq:y}. Under the white Gaussian noise model, the Fisher information matrix (FIM) is given by
\begin{equation}
\mathbf{J}(\boldsymbol{\xi};\mathbf{z},\mathbf{R}_x)=\frac{2}{\sigma^2}\operatorname{Re}\Big\{\frac{\partial\boldsymbol{\mu}^H}{\partial\boldsymbol{\xi}}\frac{\partial\boldsymbol{\mu}}{\partial\boldsymbol{\xi}^\top}\Big\}
=\begin{bmatrix}\mathbf{J}_{pp}&\mathbf{J}_{p\nu}\\ \mathbf{J}_{p\nu}^\top&\mathbf{J}_{\nu\nu}\end{bmatrix},
\label{eq:fim}
\end{equation}
where $\boldsymbol{\nu}$ collects the nuisance entries of $\boldsymbol{\xi}$ under M-II or M-III, while $\mathbf{J}$ reduces to $\mathbf{J}_{pp}$ under M-I. For fixed $\mathbf{z}$, every block of $\mathbf{J}$ is linear in $\mathbf{X}\mathbf{X}^H=T\mathbf{R}_x$ (Lemma~\ref{lem:efim}). The equivalent FIM (EFIM) for the target positions $\mathbf{p}$ is given by the generalized Schur complement
\begin{equation}
\mathbf{J}_{\mathrm{e}}(\mathbf{p};\mathbf{z},\mathbf{R}_x)=\mathbf{J}_{pp}-\mathbf{J}_{p\nu}\mathbf{J}_{\nu\nu}^{\dagger}\mathbf{J}_{p\nu}^\top\in\mathbb{R}^{2Q\times 2Q},
\label{eq:efim}
\end{equation}
which coincides with the standard Schur complement whenever $\mathbf{J}_{\nu\nu}$ is nonsingular. Under M-II, \eqref{eq:efim} simplifies to the orthogonal projection form (Lemma~\ref{lem:efim})
\begin{equation}
\mathbf{J}_{\mathrm{e}}=\frac{2}{\sigma^2}\operatorname{Re}\big\{\tilde{\mathbf{D}}^H\mathbf{P}^\perp_{\mathbf{S}_\alpha}\tilde{\mathbf{D}}\big\},
\;\;
\mathbf{P}^\perp_{\mathbf{S}_\alpha}=\mathbf{I}-\mathbf{S}_\alpha(\mathbf{S}_\alpha^H\mathbf{S}_\alpha)^{\dagger}\mathbf{S}_\alpha^H,
\label{eq:efim_proj}
\end{equation}
where $\tilde{\mathbf{D}}=[\alpha_1\mathbf{D}_1,\dots,\alpha_Q\mathbf{D}_Q]$, $\mathbf{D}_q=\partial\mathbf{s}_q/\partial\mathbf{p}_q^\top\in\mathbb{C}^{M_RST\times 2}$, and $\mathbf{S}_\alpha=[\mathbf{s}_1,\dots,\mathbf{s}_Q]$, and the corresponding projection under M-III is given in
\eqref{eq:efim_m3_proj}. The position error bound (PEB) of target $q$ is defined as
\begin{equation}
\mathrm{PEB}_q(\mathbf{z},\mathbf{R}_x)=\sqrt{\operatorname{tr}\big([\mathbf{J}_{\mathrm{e}}^{-1}]_{q}\big)},
\label{eq:peb}
\end{equation}
where $[\cdot]_q$ extracts the $2\times2$ diagonal subblock corresponding to target $q$. When $\mathbf{J}_{\mathrm{e}}$ is singular, we set the worst-target PEB to $+\infty$. In the presence of placement errors, all quantities above are evaluated at the realized coordinates $\mathbf{z}+\boldsymbol{\delta}_z$.

\section{Problem Formulation}\label{sec:problem}
This section formulates the joint geometry and covariance design problem and highlights the structural properties exploited in the subsequent analysis and algorithm design.

\subsection{Joint Geometry and Resource Design}

The mechanical element coordinates $\mathbf{z}$ are reconfigured on a slow timescale, whereas the transmit covariances $\{\mathbf{Q}_k\}$ and $\mathbf{S}$ are adapted per CPI. We jointly optimize $(\mathbf{z},\{\mathbf{Q}_k\},\mathbf{S})$ to minimize the squared worst-target PEB subject to per-user SINR and per-AP feed-power constraints:
\begin{subequations}\label{eq:P1}
\begin{align}
(\mathrm{P1}):\;&\min_{\mathbf{z},\{\mathbf{Q}_k\},\mathbf{S}}\;
\max_{q}\;\operatorname{tr}\Big(\big[\mathbf{J}_{\mathrm{e}}^{-1}\big(\bar{\mathbf{p}};\mathbf{z},\textstyle\sum_k\mathbf{Q}_k+\mathbf{S}\big)\big]_q\Big)
\label{eq:P1obj}\\
&\;\text{s.t.}\;\;
\mathrm{SINR}_k(\mathbf{z}_T,\{\mathbf{Q}_i\},\mathbf{S})\ge\gamma_k,\;\forall k,
\label{eq:P1sinr}\\
&\hphantom{\;\text{s.t.}\;\;}\operatorname{tr}\Big(\mathbf{E}_m\big(\textstyle\sum_k\mathbf{Q}_k+\mathbf{S}\big)\Big)\le P_m,\;\forall m,
\label{eq:P1pow}\\
&\hphantom{\;\text{s.t.}\;\;}\mathbf{z}\in\mathcal{Z},
\label{eq:P1z}\\
&\hphantom{\;\text{s.t.}\;\;}\mathbf{Q}_k\succeq\mathbf{0},\;\operatorname{rank}(\mathbf{Q}_k)\le1,\;\mathbf{S}\succeq\mathbf{0}.
\label{eq:P1psd}
\end{align}
\end{subequations}
The objective \eqref{eq:P1obj} is the squared worst-target PEB. Since the true target parameters are unknown at design time, it is evaluated at nominal target positions $\bar{\mathbf{p}}=[\bar{\mathbf{p}}_1^\top,\dots,\bar{\mathbf{p}}_Q^\top]^\top$ and nominal scattering coefficients $\{\bar{\alpha}_q\}$, where the coefficients enter $\mathbf{J}_{\mathrm{e}}$ through the cross-terms between targets. The EFIM is evaluated under M-II by default, and under M-III the design uses the nominal waveguide constants $\boldsymbol{\theta}_g$. The worst-target criterion prevents well-illuminated targets from masking an unobservable one. Constraint \eqref{eq:P1sinr} guarantees the SINR target $\gamma_k$ of each user, and \eqref{eq:P1pow} limits the feed power of each Tx AP to $P_m$. Constraint \eqref{eq:P1z} enforces the ordering and minimum spacing in \eqref{eq:order} on every transmit waveguide and receive segment, and \eqref{eq:P1psd} keeps the covariances positive semidefinite and restricts each user to a single beamforming vector $\mathbf{v}_k$. Problem (P1) is nonconvex because of the rank-one constraints and because $\mathbf{z}$ enters the objective and the SINR constraints through the channel phases.

\subsection{Structure of the Problem}

Problem (P1) exhibits three key structural properties that govern both its identifiability limits and its algorithmic tractability.

\emph{1) Rapid spatial oscillation in $\mathbf{z}$:} Each element coordinate enters the channel through the phase factor $e^{-j\beta_g z}\,e^{-j2\pi\Vert\tilde{\mathbf{p}}-\boldsymbol{\psi}(z)\Vert/\lambda}$. The guided phase has period $\lambda_g=\lambda/n_{\mathrm{eff}}$, which is only a few millimeters at millimeter-wave carriers; both the objective and the SINR constraints are highly multimodal in $\mathbf{z}$, making local search steps meaningful only within a radius of $\lambda_g/2$.

\emph{2) Feed-port rank bottleneck:} By \eqref{eq:y}, every target signature and its spatial derivatives lie in a subspace of complex dimension $d=\operatorname{rank}(\mathbf{R}_x)M_RS\le M_TWM_RS$, regardless of $N$ and $N_R$. Projecting out the nuisance parameters yields the rank bound
\begin{equation}
\operatorname{rank}\mathbf{J}_{\mathrm{e}}\le\min\{2Q,\;2d-n_\nu\},
\label{eq:rankbound}
\end{equation}
where $n_\nu$ is the rank of the real nuisance Jacobian. When the target signatures are linearly independent (and the calibration derivatives under M-III are linearly independent of them), we have $n_\nu=2Q$ under M-II and $n_\nu=2Q+2$ under M-III. Consequently, a single-waveguide, single-port AP pair with a nonzero target signature gives $\mathbf{J}_{\mathrm{e}}=\mathbf{0}$ for any $N$, and whenever the right-hand side of \eqref{eq:rankbound} falls below $2Q$, the objective of (P1) is $+\infty$.

\emph{3) Tractable covariance subproblem for fixed $\mathbf{z}$:} When $\mathbf{z}$ is fixed, each target's objective term is the trace of the inverse of the generalized Schur complement $\mathbf{J}_{\mathrm{e}}$, while the full FIM $\mathbf{J}$ is linear in $\mathbf{R}_x=\sum_k\mathbf{Q}_k+\mathbf{S}$. Introducing auxiliary variables $\{t_{q,i}\}$ and $u$, the inner covariance design admits the epigraph form
\begin{subequations}\label{eq:P1in}
\begin{align}
(\mathrm{P1}\text{-in}):\;&\min_{\{\mathbf{Q}_k\},\mathbf{S},\{t_{q,i}\},u}\; u \\
&\;\text{s.t.}\;\;
\textstyle\sum_{i=1}^{2}t_{q,i}\le u,\; q=1,\dots,Q,\\
&\hphantom{\;\text{s.t.}\;\;}\begin{bmatrix}\mathbf{J}(\boldsymbol{\xi};\mathbf{z},\mathbf{R}_x) & \mathbf{e}_{q,i}\\ \mathbf{e}_{q,i}^\top & t_{q,i}\end{bmatrix}\succeq\mathbf{0},\;\; \forall q,\,i,
\label{eq:lmi}\\
&\hphantom{\;\text{s.t.}\;\;}\eqref{eq:P1sinr},\;\eqref{eq:P1pow},\;\mathbf{Q}_k\succeq\mathbf{0},\;\mathbf{S}\succeq\mathbf{0},
\end{align}
\end{subequations}
where $\mathbf{e}_{q,i}$ is the unit vector selecting the $i$-th coordinate of $\mathbf{p}_q$ in $\boldsymbol{\xi}$. Since $\mathbf{J}\succeq\mathbf{0}$, the generalized Schur complement lemma ensures that \eqref{eq:lmi} holds if and only if $\mathbf{e}_{q,i}\in\operatorname{range}(\mathbf{J})$ and $t_{q,i}\ge\mathbf{e}_{q,i}^\top\mathbf{J}^\dagger\mathbf{e}_{q,i}$. All position unit vectors lie in $\operatorname{range}(\mathbf{J})$ if and only if $\mathbf{J}_{\mathrm{e}}$ in \eqref{eq:efim} is nonsingular, in which case the position block of $\mathbf{J}^\dagger$ equals $\mathbf{J}_{\mathrm{e}}^{-1}$ even when $\mathbf{J}_{\nu\nu}$ is singular. Thus, solving (P1-in) with $\operatorname{rank}(\mathbf{Q}_k)\le 1$ is equivalent to solving (P1) at a fixed $\mathbf{z}$, and (P1-in) is infeasible exactly when the objective of (P1) is $+\infty$.

Because $\mathbf{J}$ is linear in $\{\mathbf{Q}_k\}$ and $\mathbf{S}$, the linear matrix inequality (LMI) in \eqref{eq:lmi} is affine in the covariances, and \eqref{eq:P1sinr} becomes linear after clearing the denominator. Dropping the rank-one constraints in \eqref{eq:P1psd} turns (P1-in) into a convex semidefinite program (SDP) from which a rank-one solution can be recovered without loss. This two-timescale structure motivates solving (P1) via an outer coordinate search over $\mathbf{z}$ that evaluates each candidate geometry by solving (P1-in).

\section{Feed-Port-Limited Identifiability}\label{sec:limits}

This section bounds the rank of the position EFIM \eqref{eq:efim} and identifies where its information comes from. We call the target positions identifiable when $\mathbf{J}_{\mathrm{e}}$ is nonsingular, i.e., when the worst-target PEB in \eqref{eq:peb} is finite. A singular $\mathbf{J}_{\mathrm{e}}$ means that some change of the positions either leaves the noise-free echo unchanged to first order or cannot be distinguished from a change of the nuisance parameters.
Throughout this section, the operator $\iota(\mathbf{v})=[\operatorname{Re}\mathbf{v}^\top,\operatorname{Im}\mathbf{v}^\top]^\top$ maps a complex vector to its real representation and applies columnwise to matrices, satisfying $\operatorname{Re}\{\mathbf{x}^H\mathbf{v}\}=\iota(\mathbf{x})^\top\iota(\mathbf{v})$. Let $\mathbf{G}=\partial\boldsymbol{\mu}/\partial\boldsymbol{\xi}^\top$ denote the complex Jacobian, and partition its real representation as $\iota(\mathbf{G})=[\mathbf{G}_p,\mathbf{G}_\nu]$, where $\mathbf{G}_p$ contains the $2Q$ position columns and $\mathbf{G}_\nu$ collects the nuisance columns. We denote the real nuisance subspace by $\mathcal{N}=\operatorname{range}(\mathbf{G}_\nu)\subseteq\mathbb{R}^{2M_RST}$ with dimension $n_\nu=\dim\mathcal{N}$, and let $\boldsymbol{\Pi}_{\mathcal{N}}$ and $\boldsymbol{\Pi}^\perp_{\mathcal{N}}=\mathbf{I}-\boldsymbol{\Pi}_{\mathcal{N}}$ be the orthogonal projectors onto $\mathcal{N}$ and $\mathcal{N}^\perp$, respectively.

\subsection{Structure of the EFIM}

\begin{lemma}[EFIM structure]\label{lem:efim}
For any $\mathbf{z}\in\mathcal{Z}$ and any $\mathbf{X}$, the following properties hold:
\begin{enumerate}
\item[(i)] $\mathbf{J}$ in \eqref{eq:fim} depends on $\mathbf{X}$ only through $\mathbf{X}\mathbf{X}^H=T\mathbf{R}_x$ and is real-linear in $\mathbf{R}_x$.
\item[(ii)] With the Moore--Penrose pseudoinverse in \eqref{eq:efim}, the position EFIM factorizes as
\begin{equation}
\mathbf{J}_{\mathrm{e}}=\frac{2}{\sigma^2}\,\mathbf{A}^\top\mathbf{A},\;\; \mathbf{A}=\boldsymbol{\Pi}^\perp_{\mathcal{N}}\mathbf{G}_p .
\label{eq:efim_factor}
\end{equation}
\item[(iii)] Under M-II, $\mathcal{N}=\iota(\operatorname{span}_{\mathbb{C}}\{\mathbf{s}_1,\dots,\mathbf{s}_Q\})$, and \eqref{eq:efim_factor} reduces to the complex projection form in \eqref{eq:efim_proj}.
\item[(iv)] Under M-III, let $\boldsymbol{\Psi}=[\partial\boldsymbol{\mu}/\partial\alpha_g,\partial\boldsymbol{\mu}/\partial\beta_g]$. The position EFIM $\mathbf{J}_{\mathrm{e}}^{\mathrm{III}}$ is related to the M-II EFIM $\mathbf{J}_{\mathrm{e}}^{\mathrm{II}}$ by
\begin{equation}
\mathbf{J}_{\mathrm{e}}^{\mathrm{III}}=\mathbf{J}_{\mathrm{e}}^{\mathrm{II}}-\big(\mathbf{J}^{\mathrm{II}}_{\theta p}\big)^\top\big(\mathbf{J}^{\mathrm{II}}_{\theta\theta}\big)^{\dagger}\mathbf{J}^{\mathrm{II}}_{\theta p},
\label{eq:efim_m3}
\end{equation}
where $\mathbf{J}^{\mathrm{II}}_{\theta p}=\frac{2}{\sigma^2}\operatorname{Re}\{\boldsymbol{\Psi}^H\mathbf{P}^\perp_{\mathbf{S}_\alpha}\tilde{\mathbf{D}}\}$ and $\mathbf{J}^{\mathrm{II}}_{\theta\theta}=\frac{2}{\sigma^2}\operatorname{Re}\{\boldsymbol{\Psi}^H\mathbf{P}^\perp_{\mathbf{S}_\alpha}\boldsymbol{\Psi}\}$.
\end{enumerate}
\end{lemma}
\begin{IEEEproof}
See Appendix~\ref{app:lem_efim}.
\end{IEEEproof}

Under the waveguide law in \eqref{eq:b} and \eqref{eq:Cr},
$\partial_{\beta_g}\boldsymbol{\mu}
=2j\,\partial_{\alpha_g}\boldsymbol{\mu}$.
Thus, the two real calibration derivatives span one complex direction,
and \eqref{eq:efim_m3} is equivalently
\begin{equation}
\mathbf{J}_{\mathrm{e}}^{\mathrm{III}}
=\frac{2}{\sigma^2}\operatorname{Re}
\big\{\tilde{\mathbf{D}}^H
\mathbf{P}^{\perp}_{[\mathbf{S}_\alpha,\,
\partial_{\alpha_g}\boldsymbol{\mu}]}
\tilde{\mathbf{D}}\big\}.
\label{eq:efim_m3_proj}
\end{equation}
The calibration direction removes two additional real dimensions if
$\mathbf{P}^{\perp}_{\mathbf{S}_\alpha}
\partial_{\alpha_g}\boldsymbol{\mu}\neq\mathbf{0}$.
\subsection{Rank Bound}

\begin{theorem}[Feed-port rank bound]\label{thm:rank}
Let $d=\operatorname{rank}(\mathbf{R}_x)\,M_RS$. For every $\mathbf{z}\in\mathcal{Z}$, every $N,N_R$, and every target configuration:
\begin{enumerate}
\item[(a)] Every column of $\mathbf{G}$ lies in the subspace $\mathcal{V}=\operatorname{range}(\mathbf{X}^\top)\otimes\mathbb{C}^{M_RS}$, which has complex dimension $d\le M_TW\cdot M_RS$.
\item[(b)] $\operatorname{rank}\mathbf{J}_{\mathrm{e}}\le\min\{2Q,\;2d-n_\nu\}$.
\item[(c)] If $\mathbf{J}_{\nu\nu}$ is nonsingular, then $n_\nu=2Q$ under M-II and $n_\nu=2Q+2$ under M-III, yielding $\operatorname{rank}\mathbf{J}_{\mathrm{e}}\le\min\{2Q,2(d-Q)\}$ and $\operatorname{rank}\mathbf{J}_{\mathrm{e}}\le\min\{2Q,2(d-Q)-2\}$, respectively. Moreover, if $d\le Q$ and $\operatorname{rank}\mathbf{S}_\alpha=d$, then $\mathbf{J}_{\mathrm{e}}=\mathbf{0}$ under both models.
\item[(d)] Fix $(M_T,W,M_R,S,N,N_R,Q)$ and
$r=\operatorname{rank}(\mathbf{R}_x)$. If the bound in (c) is attained at
one configuration with $\operatorname{rank}\mathbf{X}=r$ and nonsingular
$\mathbf{J}_{\nu\nu}$, then it is attained almost everywhere on
$\operatorname{relint}\mathcal{Z}\times\mathbb{R}^{2Q}
\times\mathbb{C}^{Q}\times\mathcal{M}_r$, where $\mathcal{M}_r$ is the
manifold of rank-$r$ transmit matrices and almost everywhere is
understood in local real coordinates.
\end{enumerate}
\end{theorem}

The transmit covariance rank and the number of receive ports determine
the observation dimension $d$. The rank of the nuisance Jacobian
determines how many of its $2d$ real dimensions are removed. The
simplified bounds in (c) apply when the target signatures are linearly
independent and, under M-III, the calibration direction lies outside
their span. Otherwise, the general bound in (b) applies. Part~(d)
establishes generic tightness for each setting in which a configuration
with a nonsingular nuisance block attains the bound.

\begin{corollary}[Communication-only transmission]\label{cor:commonly}
Suppose $\mathbf{S}=\mathbf{0}$ and $\mathbf{x}[t]=\sum_{k=1}^K\mathbf{v}_ks_k[t]$, so that $d=\operatorname{rank}(\mathbf{R}_x)M_RS\le KM_RS$, and assume $\operatorname{rank}\mathbf{S}_\alpha=\min\{Q,d\}$
at the configuration under consideration. Under M-II:

\begin{enumerate}
\item[(a)] If $d<2Q$ (in particular if $KM_RS<2Q$), then $\mathbf{J}_{\mathrm{e}}$ is singular and the objective of (P1) is $+\infty$.
\item[(b)] If $d\le Q$ (in particular if $KM_RS\le Q$), then $\mathbf{J}_{\mathrm{e}}=\mathbf{0}$.
\end{enumerate}
Under M-III, conclusion (a) also holds at $d=2Q$ if
$\mathbf{P}^{\perp}_{\mathbf{S}_\alpha}
\partial_{\alpha_g}\boldsymbol{\mu}\neq\mathbf{0}$.
\end{corollary}

Under M-II and the signature-rank condition, a finite worst-target PEB
requires $d\ge 2Q$. If $KM_RS<2Q$, a dedicated sensing waveform must raise
the transmit covariance rank to at least $\lceil 2Q/(M_RS)\rceil$ for a
finite worst-target PEB to be possible.

\begin{remark}[Single-stream transmission]\label{rem:rank1}
Suppose $\operatorname{rank}(\mathbf{R}_x)=1$, i.e., $\mathbf{X}=\mathbf{w}\boldsymbol{\varsigma}^\top$ with $\mathbf{w}\in\mathbb{C}^{M_TW}$ and $\boldsymbol{\varsigma}\in\mathbb{C}^{T}$, corresponding to a single common data stream with $\mathbf{S}=\mathbf{0}$. Define the illumination coefficient $c_q(\mathbf{p}_q,\mathbf{z}_T)=\mathbf{w}^\top\mathbf{h}^T_q$. Since $\mathbf{X}^\top\mathbf{h}^T_q=c_q\boldsymbol{\varsigma}$, \eqref{eq:y} yields
\begin{equation}
\mathbf{s}_q=c_q\,(\boldsymbol{\varsigma}\otimes\mathbf{h}^R_q),\;\;
\mathbf{D}_q=(\boldsymbol{\varsigma}\otimes\mathbf{h}^R_q)\,\frac{\partial c_q}{\partial\mathbf{p}_q^\top}
+c_q\,\Big(\boldsymbol{\varsigma}\otimes\frac{\partial\mathbf{h}^R_q}{\partial\mathbf{p}_q^\top}\Big).
\label{eq:rank1_D}
\end{equation}
For $c_q\neq0$, the first term in $\mathbf{D}_q$ lies in $\operatorname{span}_{\mathbb{C}}\{\mathbf{s}_q\}$ and is annihilated by $\mathbf{P}^\perp_{\mathbf{S}_\alpha}$ in \eqref{eq:efim_proj}. Because $\mathbf{P}^\perp_{\mathbf{S}_\alpha}$ is invariant to the nonzero scalars $c_q$, the M-II EFIM reduces to
\begin{equation}
\mathbf{J}_{\mathrm{e}}=\frac{2\|\boldsymbol{\varsigma}\|^2}{\sigma^2}\operatorname{Re}\big\{\mathbf{G}_R^H\mathbf{P}^\perp_{\mathbf{H}_R}\mathbf{G}_R\big\},
\label{eq:rank1_efim}
\end{equation}
where $\mathbf{G}_R=\big[\alpha_1c_1\,\partial\mathbf{h}^R_1/\partial\mathbf{p}_1^\top,\dots,\alpha_Qc_Q\,\partial\mathbf{h}^R_Q/\partial\mathbf{p}_Q^\top\big]$ and $\mathbf{H}_R=[\mathbf{h}^R_1,\dots,\mathbf{h}^R_Q]$. Two key consequences follow. First, the transmit geometry provides no directional information: $\mathbf{z}_T$ and $\mathbf{w}$ affect $\mathbf{J}_{\mathrm{e}}$ solely through the scalar gains $\{\alpha_qc_q\}$, and for $Q=1$ only through the illumination power $\vert{}\alpha_1c_1\vert{}^2$, which scales $\mathbf{J}_{\mathrm{e}}$ without altering its eigenvectors. Optimizing the Tx pinching positions under single-stream transmission thus reduces to an illumination-gain problem, extending the single-target, single-element factorization of~\cite{bozanis2026secure} to multiple targets and segmented receive ports. Second, attaining the rank bound $\min\{2Q,2(M_RS-Q)\}$ depends exclusively on the receive geometry. By contrast, when $\operatorname{rank}(\mathbf{R}_x)>1$, the transmit-side derivative terms $(\mathbf{X}^\top\partial\mathbf{h}^T_q/\partial\mathbf{p}_q^\top)\otimes\mathbf{h}^R_q$ are generally not collinear with $\mathbf{s}_q$ and survive the projection to contribute directional information. In our numerical checks, omitting the transmit-side derivatives alters $\mathbf{J}_{\mathrm{e}}$ by less than $5\times10^{-15}$ in relative Frobenius norm when $\operatorname{rank}(\mathbf{R}_x)=1$, but changes $\mathbf{J}_{\mathrm{e}}$ by $42\%$ to $76\%$ when $\mathbf{R}_x$ has full rank.
\end{remark}

\subsection{Local and Cooperative Information}\label{sec:coop}

Theorem~\ref{thm:rank} establishes dimension bounds, but it does not identify the physical origin of the localization information. We distinguish two kinds of position information. Here, a \emph{link} is one part of the observation, such as the echoes collected at one Rx AP or over one Tx--Rx AP pair. The \emph{local} information of a link is the position information that the link provides on its own, i.e., when its echo is processed with a separate unknown scattering coefficient. It is the matrix $\mathbf{J}^{\mathrm{loc}}_\ell$ in \eqref{eq:coop}. The \emph{cooperative} information is the additional position information obtained because all links share the same unknown coefficient, which can only be extracted by processing the links jointly. It is the matrix $\mathbf{J}^{\mathrm{coop}}$ in \eqref{eq:jcoop}. For a single target ($Q=1$, gain $\alpha$, $\mathbf{s}=\mathbf{s}_1$, $\mathbf{D}=\mathbf{D}_1$) under M-II, the EFIM decomposes exactly into a local component contributed by each link and a cooperative component. Specifically, we decompose the target signature into $|\mathcal{L}|$ mutually orthogonal link components, $\mathbf{s}=\sum_{\ell\in\mathcal{L}}\mathbf{s}_\ell$ and $\mathbf{D}=\sum_\ell\mathbf{D}_\ell$, satisfying $\mathbf{s}_\ell^H\mathbf{s}_{\ell'}=0$, $\mathbf{D}_\ell^H\mathbf{s}_{\ell'}=\mathbf{0}$, and $\mathbf{D}_\ell^H\mathbf{D}_{\ell'}=\mathbf{0}$ for $\ell\neq\ell'$. We consider two such orthogonal partitions: the first groups measurements by receive AP ($\mathbf{s}_\ell$ retains the entries of $\mathbf{y}$ observed at Rx AP $\ell$), which is valid for arbitrary $\mathbf{R}_x$; the second groups measurements by Tx--Rx AP pair ($\ell=(m,r)$), which is valid when $\mathbf{R}_x$ is block diagonal across transmit APs, i.e., $\mathbf{S}_{mm'}=\mathbf{0}$ and $\sum_k\mathbf{v}_{m,k}\mathbf{v}_{m',k}^H=\mathbf{0}$ for $m\neq m'$. For each link, we define its energy as $\mathcal{E}_\ell=\|\mathbf{s}_\ell\|^2$. Let $\mathcal{L}_+=\{\ell:\mathcal{E}_\ell>0\}$, which is nonempty whenever $\mathbf{s}\neq\mathbf{0}$. For every $\ell\in\mathcal{L}_+$, we define the gain gradient as $\boldsymbol{\chi}_\ell=\mathbf{D}_\ell^H\mathbf{s}_\ell/\mathcal{E}_\ell\in\mathbb{C}^2$ and compute its energy-weighted mean as $\bar{\boldsymbol{\chi}}=\sum_{\ell\in\mathcal{L}_+}\mathcal{E}_\ell\boldsymbol{\chi}_\ell/\sum_{\ell\in\mathcal{L}_+}\mathcal{E}_\ell$.

\begin{proposition}[Local--cooperative decomposition]\label{prop:coop}
Under M-II with $Q=1$ and any orthogonal link partition, the position EFIM decomposes as
\begin{equation}
\mathbf{J}_{\mathrm{e}}=\sum_{\ell\in\mathcal{L}}\mathbf{J}^{\mathrm{loc}}_\ell+\mathbf{J}^{\mathrm{coop}},
\label{eq:coop}
\end{equation}
where $\mathbf{J}^{\mathrm{loc}}_\ell=\frac{2|\alpha|^2}{\sigma^2}\operatorname{Re}\{\mathbf{D}_\ell^H\mathbf{P}^\perp_{\mathbf{s}_\ell}\mathbf{D}_\ell\}$ is the local EFIM that link $\ell$ would provide if it had an independent unknown gain ($\mathbf{P}^\perp_{\mathbf{s}_\ell}=\mathbf{I}$ if $\mathbf{s}_\ell=\mathbf{0}$), and
\begin{equation}
\mathbf{J}^{\mathrm{coop}}=\frac{2|\alpha|^2}{\sigma^2}\sum_{\ell\in\mathcal{L}_+}\mathcal{E}_\ell\operatorname{Re}\big\{(\boldsymbol{\chi}_\ell-\bar{\boldsymbol{\chi}})(\boldsymbol{\chi}_\ell-\bar{\boldsymbol{\chi}})^H\big\}.
\label{eq:jcoop}
\end{equation}
Moreover:
\begin{enumerate}
\item[(a)] $\mathbf{0}\preceq\mathbf{J}^{\mathrm{coop}}\preceq\sum_{\ell\in\mathcal{L}_+}\mathbf{J}^{\mathrm{nuis}}_\ell$, where $\mathbf{J}^{\mathrm{nuis}}_\ell=\frac{2|\alpha|^2}{\sigma^2}\mathcal{E}_\ell\operatorname{Re}\{\boldsymbol{\chi}_\ell\boldsymbol{\chi}_\ell^H\}$ is the information that link $\ell$ loses to its own gain.
\item[(b)] For any $\mathbf{v}\in\mathbb{R}^2$, $\mathbf{v}^\top\mathbf{J}^{\mathrm{coop}}\mathbf{v}=\frac{2|\alpha|^2}{\sigma^2}\sum_{\ell\in\mathcal{L}_+}\mathcal{E}_\ell|\mathbf{v}^\top(\boldsymbol{\chi}_\ell-\bar{\boldsymbol{\chi}})|^2$, which vanishes if and only if $\mathbf{v}^\top\boldsymbol{\chi}_\ell$ takes the same complex value across all links in $\mathcal{L}_+$.
\end{enumerate}
\end{proposition}

The matrix $\mathbf{J}^{\mathrm{coop}}$ captures the information recovered from all links sharing a single unknown gain. Part~(a) shows that cooperation can recover at most the sum of the information lost by the individual links, while Part~(b) shows  $\mathbf{J}^{\mathrm{coop}}$ is strictly positive along every direction in which the gain gradients vary across links. By contrast, the local term $\mathbf{J}^{\mathrm{loc}}_\ell$ vanishes whenever link $\ell$ provides only a single complex observation dimension and has a nonzero signature $\mathbf{s}_\ell$ (whereas a link with $\mathbf{s}_\ell=\mathbf{0}$ can still contribute through $\mathbf{D}_\ell$). Specializing this decomposition to individual port pairs yields the following result.

\begin{corollary}[Port-pair links]\label{cor:singleport}
Let $\mathbf{R}_x$ be diagonal, corresponding to independent streams across the $M_TW$ feed ports, and partition the observation by feed-port/output-port pairs $\ell=(f,o)$ with $f=(m,w)$ and $o=(r,s)$, which is orthogonal for any $W$ and $S$. With $\mathbf{h}^T=\mathbf{h}^T_1$ and $\mathbf{h}^R=\mathbf{h}^R_1$, each link has the scalar effective channel $a_\ell(\mathbf{p})=[\mathbf{h}^T]_f[\mathbf{h}^R]_o$, phase gradient $\boldsymbol{\phi}_\ell=\nabla_{\mathbf{p}}\arg a_\ell$, and log-amplitude gradient $\boldsymbol{\tau}_\ell=\nabla_{\mathbf{p}}\ln|a_\ell|$. Assuming $a_\ell(\mathbf{p})\neq0$ and $[\mathbf{R}_x]_{ff}>0$ for all $\ell$ so that $\boldsymbol{\phi}_\ell$ and $\boldsymbol{\tau}_\ell$ are well-defined, we have $\mathbf{J}^{\mathrm{loc}}_\ell=\mathbf{0}$, $\mathcal{E}_\ell=T[\mathbf{R}_x]_{ff}|a_\ell|^2$, $\boldsymbol{\chi}_\ell=\boldsymbol{\tau}_\ell-j\boldsymbol{\phi}_\ell$, and
\begin{equation}
\mathbf{J}_{\mathrm{e}}=\frac{2|\alpha|^2}{\sigma^2}\sum_{\ell}\mathcal{E}_\ell\Big[(\boldsymbol{\phi}_\ell-\bar{\boldsymbol{\phi}})(\boldsymbol{\phi}_\ell-\bar{\boldsymbol{\phi}})^\top+(\boldsymbol{\tau}_\ell-\bar{\boldsymbol{\tau}})(\boldsymbol{\tau}_\ell-\bar{\boldsymbol{\tau}})^\top\Big],
\label{eq:scatter}
\end{equation}
where $\bar{\boldsymbol{\phi}}$ and $\bar{\boldsymbol{\tau}}$ denote the corresponding energy-weighted means.
\end{corollary}

With independent streams across the feed ports, all position information is purely cooperative regardless of $W$ and $S$: \eqref{eq:scatter} equals the energy-weighted covariance of the per-link phase and log-amplitude gradients. For a single element per guide in the far field, $\boldsymbol{\phi}_\ell\approx-\frac{2\pi}{\lambda}(\mathbf{k}^T_m+\mathbf{k}^R_r)$, where $\mathbf{k}^T_m$ and $\mathbf{k}^R_r$ are the ground-plane projections of the unit vectors from the Tx and Rx elements to the target. Since $\|\boldsymbol{\tau}_\ell\|$ is smaller than $\|\boldsymbol{\phi}_\ell\|$ by a factor on the order of the electrical distance $2\pi r/\lambda$ (with $r$ denoting the link range), $\mathbf{J}_{\mathrm{e}}$ is dominated by the angular spread of the bistatic directions across links. Compact co-located ports view the target from nearly identical directions and yield minimal gradient spread, whereas geographically distributed ports provide a wide spread.

\subsection{Element Count and Waveguide Loss}\label{sec:loss}

\begin{proposition}[Element saturation and loss sensitivity]\label{prop:loss}
The following properties hold:
\begin{enumerate}
\item[(a)] For any waveguide with coordinates $z_1<\dots<z_{N'}$ satisfying \eqref{eq:order}, let $h^{(N)}=\sum_{n\le N}b_ng(\tilde{\mathbf{p}},\boldsymbol{\psi}_n)$ denote the effective channel of its first $N$ elements. Then, for all $N'>N$,
\begin{equation}
|h^{(N')}-h^{(N)}|\le C\,\bar g\,(1-\kappa)^{N/2}e^{-\alpha_gz_{N+1}/2},
\label{eq:tail}
\end{equation}
where $C=\sqrt{\kappa}/(1-\sqrt{1-\kappa}\,e^{-\alpha_gd_{\min}/2})$ and $\bar g=\max_n\vert{}g(\tilde{\mathbf{p}},\boldsymbol{\psi}_n)\vert{}$; the same bound holds for $\Vert{}\nabla_{\mathbf{p}}h^{(N')}-\nabla_{\mathbf{p}}h^{(N)}\Vert{}$ with $\bar g$ replaced by $\max_n\Vert{}\nabla_{\mathbf{p}}g\Vert{}$. Fix a layout of $N_{\max}$ elements on each transmit waveguide and
receive segment, with $N_{\max}$ feasible on every guide. Let
$z_{\min,N+1}$ be the smallest coordinate of the $(N+1)$-st element
across these guides. Suppose that, for every prefix length
$N\in[N_0,N_{\max}]$, the nuisance Jacobian has full column rank with
smallest singular value at least $g_0>0$ and
$\lambda_{\min}(\mathbf{J}_{\mathrm{e}})\ge\lambda_0>0$.
Then there is a constant $C'$ such that
\begin{equation}
|\mathrm{PEB}(N')-\mathrm{PEB}(N)|
\le C'(1-\kappa)^{N/2}e^{-\alpha_g z_{\min,N+1}/2}
\end{equation}
for $N_0\le N<N'\le N_{\max}$, where $\mathrm{PEB}(N)$ is the
worst-target PEB using the first $N$ elements on each guide.
\item[(b)] Under the port-pair setting of Corollary~\ref{cor:singleport} (with $a_\ell\neq0$ for all $\ell$), define for each link the participation-weighted element coordinate $\zeta_\ell=\bar z^T_f+\bar z^R_o$, where $\bar z^T_f=\sum_nz_{f,n}b_{f,n}g(\tilde{\mathbf{p}},\boldsymbol{\psi}_{f,n})/[\mathbf{h}^T]_f$ and $\bar z^R_o$ is defined analogously, and let $\bar\zeta=\sum_\ell\mathcal{E}_\ell\operatorname{Re}\zeta_\ell/\sum_\ell\mathcal{E}_\ell$. Then
\begin{align}
&\frac{\partial\mathbf{J}_{\mathrm{e}}}{\partial\alpha_g}
=-\bar\zeta\,\mathbf{J}_{\mathrm{e}}
-c_\alpha\sum_\ell\mathcal{E}_\ell(\operatorname{Re}\zeta_\ell-\bar\zeta)\operatorname{Re}\{\tilde{\boldsymbol{\chi}}_\ell\tilde{\boldsymbol{\chi}}_\ell^H\}\notag\\
&-\frac{c_\alpha}{2}\sum_\ell\mathcal{E}_\ell\operatorname{Re}\big\{(\nabla_{\mathbf{p}}\zeta_\ell)^*\tilde{\boldsymbol{\chi}}_\ell^H+\tilde{\boldsymbol{\chi}}_\ell(\nabla_{\mathbf{p}}\zeta_\ell)^\top\big\},
\label{eq:dJdalpha}
\end{align}
where $c_\alpha=2\vert{}\alpha\vert{}^2/\sigma^2$ and $\tilde{\boldsymbol{\chi}}_\ell=\boldsymbol{\chi}_\ell-\bar{\boldsymbol{\chi}}$.
\end{enumerate}
\end{proposition}

Adding elements leaves the feed-port rank ceiling unchanged, though it can affect the attained rank and the conditioning of $\mathbf{J}_{\mathrm{e}}$. Part~(a) quantifies the latter effect for a fixed layout prefix: appending elements beyond the first $N$ on each waveguide alters the target signatures only by a geometrically decaying tail governed primarily by the factor $\sqrt{1-\kappa}$ per element. Within a relative tolerance $\epsilon_{\mathrm{t}}$, approximately $2\ln(1/\epsilon_{\mathrm{t}})/\lvert\ln(1-\kappa)\rvert$ elements capture the effective response, regardless of how many additional elements are installed. In Part~(b), the first term in \eqref{eq:dJdalpha} represents a uniform power attenuation that alone yields $\partial\ln\mathrm{PEB}^2/\partial\alpha_g=\bar\zeta$, which is the energy-weighted guided distance to the effective radiation center; the second term re-weights the links, and the third perturbs the effective aperture. Although \eqref{eq:dJdalpha} alone does not fix the relative magnitude of these three terms, the uniform power-loss term dominates in our numerical evaluations, accounting for a median of $102\%$ of $\partial\mathrm{PEB}^2/\partial\alpha_g$ (with the slight excess offset by a negative link re-weighting term), while the aperture term remains below $15\%$ in magnitude across $90\%$ of the tested scenarios.
\section{Joint Design and Two-Timescale Algorithm}\label{sec:algorithm}
This section solves (P1) via a two-timescale framework in which an outer search over $\mathbf{z}$ scores each candidate by the optimal value of (P1-in). We first obtain an exact, lossless SDP solution for the inner problem, then present a monotone coordinate-wise outer search matched to the oscillation scale in item~(i) of Section~\ref{sec:problem} along with its convergence guarantees, and finally analyze the computational complexity.

\subsection{Inner Problem: Exact Solution}

Every column of the complex Jacobian $\mathbf{G}$ takes the form $\mathbf{g}_i=\operatorname{vec}(\mathbf{M}_i\mathbf{X})$ for a matrix $\mathbf{M}_i\in\mathbb{C}^{M_RS\times M_TW}$ that depends on $(\mathbf{z},\bar{\mathbf{p}},\bar{\boldsymbol{\alpha}})$ (and $\boldsymbol{\theta}_g$ under M-III) but not on $\mathbf{X}$. For example, $\mathbf{M}_i=\mathbf{h}^R_q\mathbf{h}^{T\top}_q$ for $\operatorname{Re}\alpha_q$, and $\mathbf{M}_i=\alpha_q\big(\partial_{x_q}\mathbf{h}^R_q\,\mathbf{h}^{T\top}_q+\mathbf{h}^R_q\,\partial_{x_q}\mathbf{h}^{T\top}_q\big)$ for $x_q$. Consequently, each entry of the FIM is an explicit linear function of $\mathbf{R}_x=\sum_k\mathbf{Q}_k+\mathbf{S}$:
\begin{equation}
[\mathbf{J}]_{ij}=\frac{2T}{\sigma^2}\operatorname{Re}\operatorname{tr}\big(\mathbf{M}_i^H\mathbf{M}_j\mathbf{R}_x\big).
\label{eq:fim_linear}
\end{equation}
Dropping the rank-one constraints in \eqref{eq:P1psd} reduces (P1-in) to a convex SDP. In implementation, we replace the scalar LMIs in \eqref{eq:lmi} with the equivalent $2\times 2$ block form for each target: $\begin{bmatrix}\mathbf{J}&\mathbf{U}_q\\\mathbf{U}_q^\top&\boldsymbol{\Upsilon}_q\end{bmatrix}\succeq\mathbf{0}$ and $\operatorname{tr}(\boldsymbol{\Upsilon}_q)\le u$, where $\mathbf{U}_q=[\mathbf{e}_{q,1},\mathbf{e}_{q,2}]$ and $\boldsymbol{\Upsilon}_q\in\mathbb{R}^{2\times 2}$. Because the position and scattering-gain derivatives differ widely in physical scale, $\mathbf{J}$ is often ill-conditioned, with condition numbers of $10^6$ or higher. To improve numerical stability, we supply the solver with the equilibrated matrix $\boldsymbol{\Gamma}\mathbf{J}\boldsymbol{\Gamma}$ and scaled selectors $\boldsymbol{\Gamma}\mathbf{U}_q$, where $\boldsymbol{\Gamma}=\operatorname{diag}(\mathbf{J}_{\mathrm{ref}})^{-1/2}$ is formed from the FIM $\mathbf{J}_{\mathrm{ref}}$ evaluated at an isotropic reference covariance. This scaling preserves both the feasible set and the optimal value, since the scaled LMI is congruent to the original one via the nonsingular matrix $\mathrm{blkdiag}(\boldsymbol{\Gamma},\mathbf{I}_2)$, which maintains positive semidefiniteness even when $\mathbf{J}$ is singular; the true objective is then re-evaluated at the returned covariance $\mathbf{R}_x$. Moreover, this SDP relaxation is tight.

\begin{lemma}[Lossless rank-one recovery]\label{lem:rank1}
Let $(\{\mathbf{Q}_k\},\mathbf{S})$ solve the SDP relaxation of (P1-in). Define
\begin{equation}
\tilde{\mathbf{Q}}_k=\frac{\mathbf{Q}_k\mathbf{h}_k^*\mathbf{h}_k^\top\mathbf{Q}_k}{\mathbf{h}_k^\top\mathbf{Q}_k\mathbf{h}_k^*},\;\;
\tilde{\mathbf{S}}=\mathbf{S}+\sum_{k=1}^{K}\big(\mathbf{Q}_k-\tilde{\mathbf{Q}}_k\big).
\label{eq:rank1_rec}
\end{equation}
Then $\operatorname{rank}\tilde{\mathbf{Q}}_k=1$, $\tilde{\mathbf{S}}\succeq\mathbf{0}$, and $(\{\tilde{\mathbf{Q}}_k\},\tilde{\mathbf{S}})$ attains the same objective value, user SINRs, and per-AP feed powers. Hence $(\{\tilde{\mathbf{Q}}_k\},\tilde{\mathbf{S}})$ is globally optimal for (P1-in) with the rank-one constraints.
\end{lemma}

This rank-one reconstruction follows~\cite{liu2022crb} and holds here for the FIM depends on the transmit covariances solely through their sum $\mathbf{R}_x$, which \eqref{eq:rank1_rec} leaves invariant. Across $200$ random instances ($M_T=M_R=4$, $W=S=1$, $N=N_R=4$, $K=Q=2$, $\gamma_k=10$~dB), the interior-point solver directly returned rank-one matrices $\{\mathbf{Q}_k\}$ in $30$ instances, while \eqref{eq:rank1_rec} recovered a feasible rank-one solution in all $200$ instances with a relative objective discrepancy of at most $4\times10^{-13}$.

\subsection{Outer Search over the Pinching Coordinates}

Let $f(\mathbf{z})$ denote the optimal value of (P1-in) at $\mathbf{z}$, with $f(\mathbf{z})=+\infty$ when (P1-in) is infeasible. Because $f(\mathbf{z})$ oscillates along each coordinate with a period on the order of the guided wavelength $\lambda_g=\lambda/n_{\mathrm{eff}}$, local steps are meaningful only within a trust region of radius below $\lambda_g/2$. Algorithm~\ref{alg:outer} updates one coordinate at a time by evaluating at most six candidates: four local stencil points $z_i\pm\Delta$ and $z_i\pm 2\Delta$ around the current coordinate (with $2\Delta<\lambda_g/2$), together with the two endpoints of the coordinate's feasible interval to allow jumps across phase basins. A candidate is accepted only if it reduces $f$ by at least a relative fraction $\epsilon_{\mathrm{a}}$.

\begin{algorithm}[t]
\caption{Two-timescale design of $(\mathbf{z},\{\mathbf{Q}_k\},\mathbf{S})$}\label{alg:outer}
\begin{algorithmic}[1]
\REQUIRE feasible $\mathbf{z}^{(0)}\in\mathcal{Z}$ with $f(\mathbf{z}^{(0)})<\infty$; $\Delta_0<\lambda_g/4$; $\Delta_{\min}$; $\epsilon_{\mathrm{a}}\in(0,1)$
\STATE $\mathbf{z}\leftarrow\mathbf{z}^{(0)}$, $f_{\mathrm{cur}}\leftarrow f(\mathbf{z})$, $\Delta\leftarrow\Delta_0$
\WHILE{$\Delta\ge\Delta_{\min}$}
\STATE improved $\leftarrow$ false
\FOR{each coordinate $z_i$ of $\mathbf{z}$}
\STATE $[\ell_i,u_i]\leftarrow$ feasible interval of $z_i$ given the other coordinates, from \eqref{eq:order}
\STATE $\mathcal{C}\leftarrow\big(\{z_i\pm\Delta,z_i\pm2\Delta\}\cup\{\ell_i,u_i\}\big)\cap[\ell_i,u_i]$
\STATE $c^\star\leftarrow\arg\min_{c\in\mathcal{C}}f(\mathbf{z}\,\vert{}\,z_i=c)$ \hfill\COMMENT{one SDP per candidate}
\IF{$f(\mathbf{z}\,\vert{}\,z_i=c^\star)<(1-\epsilon_{\mathrm{a}})f_{\mathrm{cur}}$}
\STATE $z_i\leftarrow c^\star$, update $f_{\mathrm{cur}}$, improved $\leftarrow$ true
\ENDIF
\ENDFOR
\STATE \textbf{if} not improved \textbf{then} $\Delta\leftarrow\Delta/2$
\ENDWHILE
\STATE solve (P1-in) at $\mathbf{z}$ and apply \eqref{eq:rank1_rec}
\ENSURE $\mathbf{z},\{\tilde{\mathbf{Q}}_k\},\tilde{\mathbf{S}}$
\end{algorithmic}
\end{algorithm}

\begin{proposition}[Properties of Algorithm~\ref{alg:outer}]\label{prop:alg}
The following properties hold:
\begin{enumerate}
\item[(a)] Every iterate $\mathbf{z}$ lies in $\mathcal{Z}$ with $f(\mathbf{z})<\infty$, and the returned $(\{\tilde{\mathbf{Q}}_k\},\tilde{\mathbf{S}})$ is feasible for (P1) at the final $\mathbf{z}$.
\item[(b)] The objective sequence $f_{\mathrm{cur}}$ is monotonically non-increasing.
\item[(c)] Algorithm~\ref{alg:outer} terminates in finitely many steps. Specifically, the number of accepted moves is bounded by $\ln\!\big(f(\mathbf{z}^{(0)})/f_{\inf}\big)/\lvert\ln(1-\epsilon_{\mathrm{a}})\rvert$, where $f_{\inf}=\inf_{\mathbf{z}\in\mathcal{Z}}f(\mathbf{z})>0$, and the total number of sweeps is at most this bound plus $\lceil\log_2(\Delta_0/\Delta_{\min})\rceil+1$, with each sweep solving at most $6(M_TWN+M_RSN_R)$ SDPs.
\item[(d)] At termination, neither a single-coordinate move within the final stencil nor a jump to an interval endpoint reduces $f$ by more than the fraction $\epsilon_{\mathrm{a}}$, while the inner covariance variables are globally optimal at every visited $\mathbf{z}$.
\end{enumerate}
\end{proposition}
This guarantee does not claim global optimality in $\mathbf{z}$: the objective landscape is inherently multimodal, and the endpoint candidates merely enlarge the set of phase basins accessible within a single sweep. Across five random instances ($N=N_R=4$, eight sweeps), the first sweep already captured $10$ to $12$~dB of the total $12$ to $15$~dB PEB reduction.

\subsection{Complexity Analysis}
For a fixed geometry $\mathbf{z}$, the inner SDP involves $K+1$ Hermitian matrix variables of size $M_TW\times M_TW$, $Q$ LMIs of dimension $n_\xi+2$ (where $n_\xi=\dim\boldsymbol{\xi}$ equals $4Q$ under M-II and $4Q+2$ under M-III), and $K+M_T$ linear scalar constraints. Crucially, the SDP dimension is independent of $N$ and $N_R$: the element counts enter only when assembling the coefficient matrices $\mathbf{M}_i$, which requires $\mathcal{O}\big((M_TWN+M_RSN_R)\,Q\big)$ channel evaluations. With at most $6(M_TWN+M_RSN_R)$ SDP evaluations per outer sweep, solving a single SDP for $M_T=M_R=4$, $W=S=1$, and $K=Q=2$ takes a median of $0.07$~s using the Clarabel interior-point solver in Python, resulting in a per-sweep runtime of approximately $13$~s when $N=N_R=4$.
\begin{figure*}[t]
\centering
\includegraphics[width=0.95\textwidth]{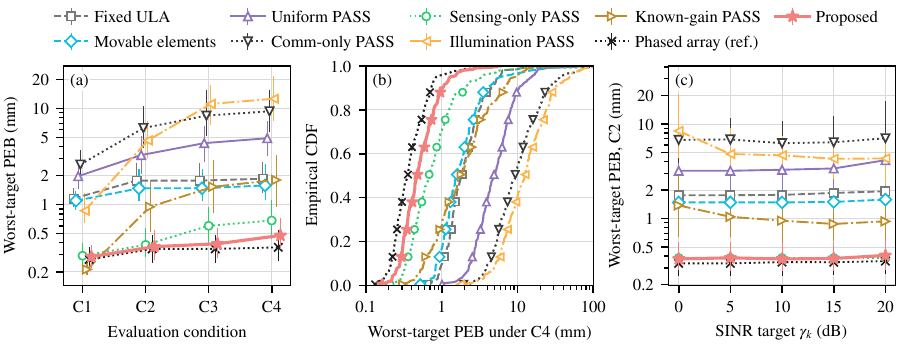}
\vspace{-10pt}
\caption{Worst-target PEB over $200$ random scenarios ($M_T=M_R=4$, $W=S=1$, $N=N_R=8$, $K=Q=2$; bars show interquartile ranges). (a) Median PEB across C1--C4. (b) Empirical CDF under C4 ($\sigma_\delta=\lambda_g/20$, $20$ draws). (c) Median PEB under C2 versus the SINR target $\gamma_k$ (infeasible instances set to $\mathrm{PEB}=\infty$; sensing-only has no SINR constraint).}
\label{fig:core}
\end{figure*}

\section{Numerical Results}\label{sec:numerical}

This section evaluates the proposed design against equal-hardware and phased-array benchmarks across model mismatches, placement errors, and varying SINR targets, and verifies our analytical findings on feed-port rank, element-count scaling, and distributed versus co-located AP deployments.

\subsection{Setup and Benchmarks}\label{sec:setup}

\begin{figure*}[t]
\centering
\includegraphics[width=0.95\textwidth]{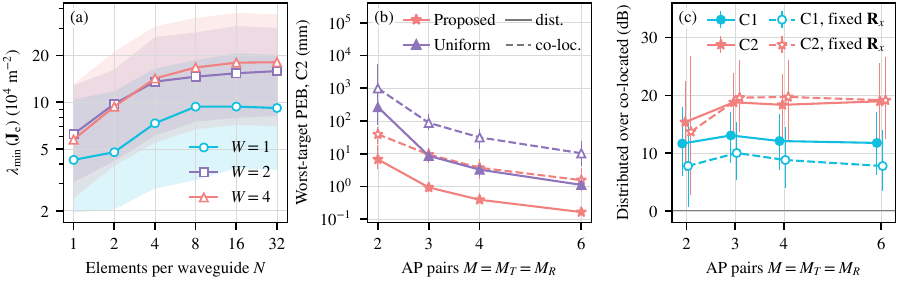}
\vspace{-10pt}
\caption{(a) Median $\lambda_{\min}(\mathbf{J}_{\mathrm{e}})$ versus $N$ for $W$ waveguides per Tx AP (M-II, $S=1$, $N_R=8$, isotropic covariance, first $N$ elements of a fixed layout). (b) Median worst-target PEB under C2 versus AP pairs $M$ for distributed ($W=S=1$) and co-located ($W=S=M$, $0.5$~m spacing) deployments under equal hardware and total power (infeasible instances set to $\mathrm{PEB}=\infty$). (c) Median paired gain of distributed over co-located deployment under C1 and C2 for the proposed design (solid) and uniform elements with $\mathbf{R}_x=P_m\mathbf{I}$ and no SINR constraint (dashed). Bands and bars show interquartile ranges.}
\label{fig:struct}
\end{figure*}

Unless stated otherwise, $M_T=M_R=4$ APs are placed uniformly at random in $[-15,15]^2$~m at height $H=5$~m, with $W=S=1$, $L=3$~m, $L_R=1.5$~m, and $N=N_R=8$. We set $f_c=28$~GHz ($\lambda=10.7$~mm), $n_{\mathrm{eff}}=1.4$, $\kappa=0.3$, $\alpha_g=0.1$~Np/m, and $d_{\min}=\lambda/2$. Furthermore, $K=2$ users and $Q=2$ targets are placed uniformly in $[-10,10]^2$~m with $\vert{}\alpha_q\vert{}=1$ and uniform phase, with $T=64$, $\sigma^2=10^{-2}$, $\sigma_k^2=10^{-4}$, $P_m=1$, $\gamma_k=10$~dB, and $\eta_0=1$ in \eqref{eq:g}. Under this normalized link budget, the proposed design yields a median user SNR of $28$~dB (under coherent full-power transmission) and a median integrated target echo SNR of $20$~dB. Unless stated otherwise, designs assume oracle knowledge of $\boldsymbol{\alpha}$ and $\boldsymbol{\theta}_g$. Algorithm~\ref{alg:outer} uses $\Delta_0=\lambda_g/5$, $\Delta_{\min}=\Delta_0/32$, $\epsilon_{\mathrm{a}}=10^{-4}$, and terminates after four sweeps (two in Figs.~\ref{fig:struct}(b) and~(c)); all geometry-searched benchmarks share this same search routine and stopping rule. Over $200$ shared random scenarios, we report medians, interquartile ranges, and the median paired gain $20\log_{10}(\mathrm{PEB}_{\text{scheme}}/\mathrm{PEB}_{\text{proposed}})$ in terms of the worst-target PEB. 

All schemes share the same numbers of transmit RF chains and receive ports and the same per-AP feed power, so waveguide loss is charged only to PASS. We compare nine benchmark and reference schemes:
\begin{itemize}
\item \emph{Fixed ULA:} $W$ ($S$) lossless $\lambda/2$-spaced antennas per Tx (Rx) AP, each with a dedicated RF chain, and covariances from (P1-in).
\item \emph{Movable elements}~\cite{wong2021fluid,zhu2024movable,he2026crb}: fixed ULA with each antenna movable by up to $\lambda$ along the array normal, jointly optimized via the proposed objective and search.
\item \emph{Uniform PASS:} evenly spaced elements on each waveguide with covariances from (P1-in).
\item \emph{Comm-only PASS:} sets $\mathbf{S}=\mathbf{0}$; positions and $\{\mathbf{Q}_k\}$ minimize total feed power under SINR constraints, followed by uniform power scaling until one AP saturates.
\item \emph{Sensing-only PASS:} solves (P1) without SINR constraints.
\item \emph{Illumination PASS}~\cite{mao2026multi}: maximizes $\min_q\mathbf{h}^{T\top}_q\mathbf{R}_x\mathbf{h}^{T*}_q$ under the constraints of (P1).
\item \emph{Known-gain PASS}~\cite{geng2026swan}: solves (P1) under M-I.
\item \emph{Phased-array reference:} uniform-PASS positions on apertures $L$ and $L_R$ fed by an ideal lossless corporate network with $M_TWN+M_RSN_R$ phase shifters, replacing \eqref{eq:b} with $e^{j\theta_n}/\sqrt{N}$ (Tx) and $e^{j\theta_n}/\sqrt{N_R}$ (Rx), optimized via the proposed objective and search.
\item \emph{Proposed design:} Algorithm~\ref{alg:outer} under M-II, initialized from the best (under M-II) among random coordinates, uniform coordinates, and the four searched PASS benchmarks.
\end{itemize}

Each design is evaluated under four conditions: C1 (M-I, exact positions), C2 (M-II, exact positions; main model), C3 (M-III, exact positions), and C4 (M-III with realized positions $\mathbf{z}+\boldsymbol{\delta}_z$). In C4, $\boldsymbol{\delta}_z$ has i.i.d.\ $\mathcal{N}(0,\sigma_\delta^2)$ entries with $\sigma_\delta=\lambda_g/20=0.38$~mm, keeping the covariances designed for $\mathbf{z}$; we report the conditional PEB (assuming known realized positions at the estimator) averaged over $20$ draws of $\boldsymbol{\delta}_z$, and evaluate the perturbed SINR separately. The proposed scheme is redesigned under M-III for C3 and C4, whereas all benchmarks are designed once. Since the fixed ULA, movable array, and phased array have no waveguides, C3 and C4 reduce to the M-II parameter set for them.

\subsection{Comparison with Benchmarks}\label{sec:core}

\begin{figure*}[t]
\centering
\includegraphics[width=0.95\textwidth]{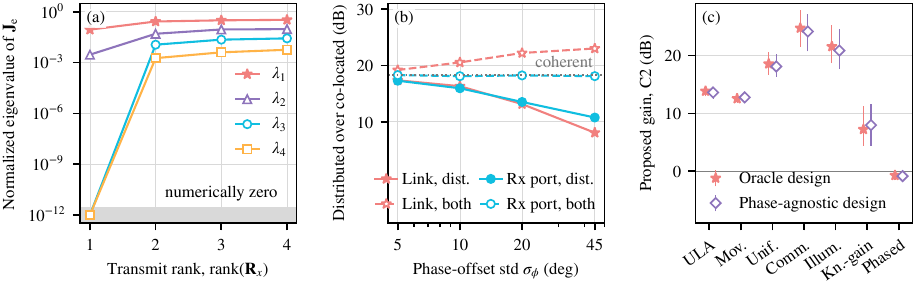}
\vspace{-10pt}
\caption{(a) Median eigenvalues of $\mathbf{J}_{\mathrm{e}}$ (normalized by $\lambda_{\max}(\mathbf{J}_{pp})$) versus transmit rank (M-II, $M_T=4$, $M_RS=3$, $Q=2$, uniform coordinates, random $\mathbf{R}_x$; floor at $10^{-12}$). (b) Median paired gain of distributed over co-located deployment ($M=4$, proposed designs from Fig.~\ref{fig:struct}) versus phase-offset standard deviation per target-link pair or per Rx port, applied to distributed links (dist.) or both deployments (both). (c) Median paired gain (with interquartile bars) of the proposed design over each benchmark under C2 for oracle and phase-agnostic ($\bar\alpha_q=|\alpha_q|$) designs.}
\label{fig:r2}
\end{figure*}

Under the main model C2 (Fig.~\ref{fig:core}(a)), the proposed design achieves the lowest PEB among all equal-hardware schemes that satisfy the SINR constraints, yielding median gains ranging from $7.2$~dB over known-gain PASS to $24.8$~dB over comm-only PASS. Uniform PASS performs worse than the fixed ULA, confirming that pinching elements are beneficial only when properly positioned. Only the phased-array reference, which requires a dedicated phase shifter per element, achieves a lower PEB (by $0.7$~dB). Moving from C1 to C2 degrades known-gain and illumination PASS by $13$ to $14$~dB, compared with only $1.6$~dB for the proposed design, because ignoring the projection in \eqref{eq:efim_proj} places the aperture where the unknown scattering gain absorbs the directional information. Meanwhile, the sensing-only reference attains nearly the same PEB as the proposed design, indicating that the $10$~dB SINR constraint incurs negligible localization loss. In Fig.~\ref{fig:r2}(c), designing each scheme with phase-agnostic nominal gains $\bar\alpha_q=|\alpha_q|$ and evaluating it under the true coefficients degrades the proposed design by a median of $0.6$~dB (and no benchmark by more), altering the relative gains under C2 by less than $1$~dB.

These gains persist under uncalibrated waveguide constants and placement errors (C3 and C4), holding across the entire distribution in Fig.~\ref{fig:core}(b). Placement errors, however, degrade communication SINR: because nominal designs leave zero SINR margin, the proposed design satisfies the SINR target in only $11\%$ of the perturbed realizations, compared with $21\%$ for the fixed ULA and $24\%$ for the phased array. Guaranteeing $90\%$ SINR satisfaction requires a back-off margin of about $7$~dB for the proposed and uniform PASS designs, versus less than $3$~dB for the fixed ULA and phased array, as placement perturbations also rotate the guided phase ($\beta_g\sigma_\delta=2\pi/20$).

Fig.~\ref{fig:core}(c) varies the SINR target $\gamma_k$. To reduce runtime across this sweep, the search for the proposed design initializes from the better of the random and uniform coordinates (which incurs a $0.3$~dB loss at $\gamma_k=10$~dB) and records $\mathrm{PEB}=\infty$ when (P1-in) is infeasible at both starting points. An infeasible initialization does not imply physical infeasibility: at $\gamma_k=20$~dB, this occurs in four scenarios, in three of which another PASS benchmark remains feasible. The PEB of the proposed design remains nearly flat up to $\gamma_k=15$~dB and increases by only $0.7$~dB at $20$~dB, outperforming every equal-hardware SINR-feasible benchmark across the entire range.

\subsection{Feed Ports, Element Count and AP Distribution}\label{sec:struct}

Across $768$ instances varying $W$, $S$, $N=N_R$, $Q$, $\operatorname{rank}(\mathbf{R}_x)$, and the parameter model, the numerical rank of $\mathbf{J}_{\mathrm{e}}$ matched the feed-port rank bound in \eqref{eq:rankbound} and remained invariant to $N$. Fig.~\ref{fig:r2}(a) illustrates the transmit-rank threshold for $M_T=4$, $M_RS=3$, and $Q=2$: under single-stream transmission ($d=3$), two eigenvalues of $\mathbf{J}_{\mathrm{e}}$ vanish as predicted by $\min\{4,2d-n_\nu\}=2$, whereas $\mathbf{J}_{\mathrm{e}}$ becomes nonsingular for transmit ranks of two or higher (and for $M_RS=2$, a single stream yields $\mathbf{J}_{\mathrm{e}}=\mathbf{0}$). In Fig.~\ref{fig:struct}(a), $\lambda_{\min}(\mathbf{J}_{\mathrm{e}})$ saturates beyond roughly eight elements per waveguide, while adding a second waveguide per AP raises it by $73\%$. In single-target tests, the PEB converges in $N$ at the geometric rate $(1-\kappa)^{N/2}$ of \eqref{eq:tail}.

Figs.~\ref{fig:struct}(b) and~(c) compare $M$ distributed AP pairs ($W=S=1$) against a single co-located pair ($W=S=M$) under identical RF chains, receive ports, elements, and total feed power. Joint designs are feasible across all $200$ scenarios for $M\ge3$; for $M=2$, the search fails in $16$ (distributed) and $3$ (co-located) scenarios, so paired gains are evaluated over mutually feasible scenarios. In a companion single-target check with diagonal covariance, \eqref{eq:scatter} matched the exact EFIM to within $5\times10^{-7}$ and exhibited the same C1-versus-C2 split ($6$ to $7$~dB under C1 versus $15$ to $17$~dB under C2). The distributed deployment reduces the median PEB of the proposed design by $18$ to $19$~dB for $M\ge3$ and by $15$~dB for $M=2$. To isolate geometry from power constraints and covariance optimization, Fig.~\ref{fig:struct}(c) repeats the comparison using uniform elements, fixed isotropic covariance $\mathbf{R}_x=P_m\mathbf{I}$, and no SINR constraint. Under C2, the distributed advantage remains $14$ to $20$~dB, confirming its geometric origin. Under known coefficients (C1), the gain drops to $8$ to $10$~dB with fixed covariance and $12$ to $13$~dB with the joint design; the extra gain under C2 stems directly from the unknown scattering coefficients, consistent with the cooperative recovery mechanism in \eqref{eq:coop} and \eqref{eq:scatter}.

Two sensitivity checks at $M=4$ qualify this gain. First, widening the co-located deployment to waveguide spacings of $2$ to $5$~m and a Tx--Rx separation of $5$ to $10$~m narrows the distributed advantage to $9$ to $11$~dB. Second, Fig.~\ref{fig:r2}(b) relaxes the common-gain assumption of Section~\ref{sec:sensing} via two zero-mean Gaussian phase-offset priors: per-target per-link offsets (scattering decorrelation) and per-Rx-port offsets (receive calibration error). When applied solely to the distributed links while keeping the co-located ports coherent, both reduce the advantage gradually to $13$ to $14$~dB at a $20^\circ$ standard deviation. When applied to both deployments, the relative advantage stays at $18$~dB or higher because the compact deployment degrades at least as fast, though the absolute PEB of both deployments worsens. Without a prior, assigning an uncoupled unknown coefficient per Rx port or per link removes most localization information in both deployments, confirming that the reported accuracy relies on inter-port phase coherence.

\section{Conclusion}\label{sec:conclusion}

This paper studied device-free target localization in pinching-antenna cell-free ISAC with unknown scattering coefficients. We bounded the rank of the position EFIM, decomposed the single-target EFIM into local and cooperative components, and developed a two-timescale covariance and placement design. Three insights were obtained. The feed ports, rather than the pinching elements, set the available observation dimensions, so communication-only transmission could not localize the targets when the transmit rank was too low. With independent feed streams, the position information was entirely cooperative, which explained why distributed ports outperformed the tested co-located ones. The joint design lowered the worst-target PEB relative to the equal-hardware benchmarks, whereas element placement errors compromised SINR feasibility.

\appendices

\section{Proof of Lemma~\ref{lem:efim}}\label{app:lem_efim}
By \eqref{eq:y}, $\boldsymbol{\mu}=\sum_q\alpha_q(\mathbf{X}^\top\mathbf{h}^T_q)\otimes\mathbf{h}^R_q$ with $\mathbf{h}^T_q,\mathbf{h}^R_q$ independent of $\mathbf{X}$, so every column of $\mathbf{G}$ is a finite sum $\sum_k(\mathbf{X}^\top\mathbf{a}_k)\otimes\mathbf{b}_k$ with $\mathbf{a}_k,\mathbf{b}_k$ independent of $\mathbf{X}$. Since $\big((\mathbf{X}^\top\mathbf{a})\otimes\mathbf{b}\big)^H\big((\mathbf{X}^\top\mathbf{a}')\otimes\mathbf{b}'\big)=T\big(\mathbf{a}^H\mathbf{R}_x^*\mathbf{a}'\big)\big(\mathbf{b}^H\mathbf{b}'\big)$, every entry of $\mathbf{G}^H\mathbf{G}$, and hence of $\mathbf{J}=\frac{2}{\sigma^2}\operatorname{Re}\{\mathbf{G}^H\mathbf{G}\}$, is real-linear in $\mathbf{R}_x$.

Since $\operatorname{Re}\{\mathbf{x}^H\mathbf{v}\}=\iota(\mathbf{x})^\top\iota(\mathbf{v})$, $\mathbf{J}=\frac{2}{\sigma^2}[\mathbf{G}_p,\mathbf{G}_\nu]^\top[\mathbf{G}_p,\mathbf{G}_\nu]$. With $\mathbf{G}_\nu(\mathbf{G}_\nu^\top\mathbf{G}_\nu)^\dagger\mathbf{G}_\nu^\top=\boldsymbol{\Pi}_{\mathcal{N}}$, \eqref{eq:efim} becomes $\frac{2}{\sigma^2}\mathbf{G}_p^\top\boldsymbol{\Pi}^\perp_{\mathcal{N}}\mathbf{G}_p$, which is \eqref{eq:efim_factor} because $\boldsymbol{\Pi}^\perp_{\mathcal{N}}$ is symmetric and idempotent.

Under M-II, the nuisance columns are $\iota(\mathbf{s}_q)$ and $\iota(j\mathbf{s}_q)$, $q=1,\dots,Q$. Since $\iota(c\,\mathbf{s})=\operatorname{Re}(c)\,\iota(\mathbf{s})+\operatorname{Im}(c)\,\iota(j\mathbf{s})$ for $c\in\mathbb{C}$, their real span is $\iota(\mathcal{S})$ with $\mathcal{S}=\operatorname{span}_{\mathbb{C}}\{\mathbf{s}_q\}$. Since $\iota(\mathbf{v})\perp\iota(\mathcal{S})$ means $\operatorname{Re}\{\mathbf{x}^H\mathbf{v}\}=0$ for both $\mathbf{x}$ and $j\mathbf{x}$, i.e., $\mathbf{x}^H\mathbf{v}=0$ for all $\mathbf{x}\in\mathcal{S}$, we have $\iota(\mathcal{S})^\perp=\iota(\mathcal{S}^\perp)$ and $\boldsymbol{\Pi}^\perp_{\mathcal{N}}\iota(\mathbf{v})=\iota(\mathbf{P}^\perp_{\mathbf{S}_\alpha}\mathbf{v})$. The position columns are $\iota(\alpha_q\mathbf{D}_q)$, so $\mathbf{A}=\iota(\mathbf{P}^\perp_{\mathbf{S}_\alpha}\tilde{\mathbf{D}})$ and $\mathbf{A}^\top\mathbf{A}=\operatorname{Re}\{\tilde{\mathbf{D}}^H\mathbf{P}^\perp_{\mathbf{S}_\alpha}\tilde{\mathbf{D}}\}$, which is \eqref{eq:efim_proj}.

Under M-III, let $\mathcal{N}_{\mathrm{II}}$ denote the subspace $\mathcal{N}$ under M-II, so that $\mathcal{N}=\mathcal{N}_{\mathrm{II}}+\operatorname{range}\iota(\boldsymbol{\Psi})$. With $\mathbf{B}=\boldsymbol{\Pi}^\perp_{\mathcal{N}_{\mathrm{II}}}\iota(\boldsymbol{\Psi})$, the orthogonal decomposition $\mathcal{N}=\mathcal{N}_{\mathrm{II}}\oplus\operatorname{range}\mathbf{B}$ gives $\boldsymbol{\Pi}^\perp_{\mathcal{N}}=\boldsymbol{\Pi}^\perp_{\mathcal{N}_{\mathrm{II}}}-\mathbf{B}(\mathbf{B}^\top\mathbf{B})^\dagger\mathbf{B}^\top$. Substituting into \eqref{eq:efim_factor} and using the M-II projection identity above for the inner products, $\frac{2}{\sigma^2}\mathbf{B}^\top\mathbf{B}=\mathbf{J}^{\mathrm{II}}_{\theta\theta}$ and $\frac{2}{\sigma^2}\mathbf{B}^\top\boldsymbol{\Pi}^\perp_{\mathcal{N}_{\mathrm{II}}}\mathbf{G}_p=\mathbf{J}^{\mathrm{II}}_{\theta p}$, which yields \eqref{eq:efim_m3}. \hfill$\blacksquare$

\section{Proofs of Theorem~\ref{thm:rank} and Corollary~\ref{cor:commonly}}\label{app:thm_rank}

For the subspace property, note that $\mathbf{s}_q=(\mathbf{X}^\top\mathbf{h}^T_q)\otimes\mathbf{h}^R_q\in\mathcal{V}$. For any scalar parameter $\theta\in\{x_q,y_q,\alpha_g,\beta_g\}$,
\begin{equation}
\frac{\partial\mathbf{s}_q}{\partial\theta}=\Big(\mathbf{X}^\top\frac{\partial\mathbf{h}^T_q}{\partial\theta}\Big)\otimes\mathbf{h}^R_q+\big(\mathbf{X}^\top\mathbf{h}^T_q\big)\otimes\frac{\partial\mathbf{h}^R_q}{\partial\theta}\in\mathcal{V},
\end{equation}
and the derivatives with respect to $\operatorname{Re}\alpha_q,\operatorname{Im}\alpha_q$ are $\mathbf{s}_q,j\mathbf{s}_q\in\mathcal{V}$. Hence every column of $\mathbf{G}$ lies in $\mathcal{V}$. Since $\operatorname{rank}\mathbf{X}^\top=\operatorname{rank}\mathbf{X}\mathbf{X}^H=\operatorname{rank}\mathbf{R}_x$, $\dim_{\mathbb{C}}\mathcal{V}=\operatorname{rank}(\mathbf{R}_x)\,M_RS=d$. Finally $\mathbf{R}_x\in\mathbb{C}^{M_TW\times M_TW}$ by Remark~\ref{rem:feed}, so $d\le M_TW\,M_RS$.

For the rank bound, the subspace property gives $\mathcal{N}\subseteq\iota(\mathcal{V})$ and $\operatorname{range}\mathbf{G}_p\subseteq\iota(\mathcal{V})$, where $\dim_{\mathbb{R}}\iota(\mathcal{V})=2d$. For $\mathbf{v}\in\iota(\mathcal{V})$, $\boldsymbol{\Pi}^\perp_{\mathcal{N}}\mathbf{v}=\mathbf{v}-\boldsymbol{\Pi}_{\mathcal{N}}\mathbf{v}\in\iota(\mathcal{V})\cap\mathcal{N}^\perp$, a subspace of dimension $2d-n_\nu$. The matrix $\mathbf{A}$ in \eqref{eq:efim_factor} has $2Q$ columns in this subspace, so $\operatorname{rank}\mathbf{J}_{\mathrm{e}}=\operatorname{rank}\mathbf{A}\le\min\{2Q,2d-n_\nu\}$.

For the simplified bounds, $\mathbf{J}_{\nu\nu}=\frac{2}{\sigma^2}\mathbf{G}_\nu^\top\mathbf{G}_\nu$ is nonsingular if and only if $\mathbf{G}_\nu$ has full column rank, i.e., $n_\nu$ equals the number of nuisance parameters: $2Q$ under M-II and $2Q+2$ under M-III. Substituting into the rank bound gives the two bounds. If $d\le Q$ and $\operatorname{rank}\mathbf{S}_\alpha=d$, then $\operatorname{span}_{\mathbb{C}}\{\mathbf{s}_q\}=\mathcal{V}$, so $\mathcal{N}\supseteq\iota(\mathcal{V})$ under both models, $\mathbf{A}=\mathbf{0}$ and $\mathbf{J}_{\mathrm{e}}=\mathbf{0}$.

For generic tightness, the entries of $[\mathbf{G}_\nu,\mathbf{G}_p]$ are real analytic in local coordinates on $\Omega=\operatorname{relint}\mathcal{Z}\times\mathbb{R}^{2Q}\times\mathbb{C}^{Q}\times\mathcal{M}_r$, since the propagation distances are bounded below by $H$, and $\Omega$ is connected because $\mathcal{Z}$ is convex and $\mathcal{M}_r$ is connected. By \eqref{eq:efim_factor}, $\operatorname{rank}\mathbf{J}_{\mathrm{e}}=\operatorname{rank}[\mathbf{G}_\nu,\mathbf{G}_p]-\operatorname{rank}\mathbf{G}_\nu$. Let $\beta$ denote the simplified bound and $\bar n_\nu$ the number of nuisance parameters. At the stated configuration, a minor of order $\bar n_\nu+\beta$ that uses all nuisance columns is nonzero, and by continuity it remains nonzero at a nearby point of $\operatorname{relint}\mathcal{Z}$ if the layout lies on the relative boundary. This minor is a nontrivial real-analytic function on $\Omega$, so its zero set has measure zero in local coordinates~\cite{mityagin2015zero}. Wherever it is nonzero, $\mathbf{G}_\nu$ has full column rank and $\operatorname{rank}\mathbf{J}_{\mathrm{e}}\ge\beta$, while the simplified bound supplies the reverse inequality.

For Corollary~\ref{cor:commonly}, $\mathbf{S}=\mathbf{0}$ confines every $\mathbf{x}[t]$ to $\operatorname{span}\{\mathbf{v}_1,\dots,\mathbf{v}_K\}$, so $\operatorname{rank}\mathbf{R}_x\le K$ and $d\le KM_RS$ without the large-$T$ approximation. Under M-II with $\operatorname{rank}\mathbf{S}_\alpha=\min\{Q,d\}$, if $d\le Q$, then $\operatorname{rank}\mathbf{S}_\alpha=d$ and Theorem~\ref{thm:rank}(c) gives $\mathbf{J}_{\mathrm{e}}=\mathbf{0}$; if $Q<d<2Q$, then $n_\nu=2Q$ and Theorem~\ref{thm:rank}(b) gives $\operatorname{rank}\mathbf{J}_{\mathrm{e}}\le2(d-Q)<2Q$. In both cases $\mathbf{J}_{\mathrm{e}}$ is singular and the worst-target PEB is infinite. Under M-III at $d=2Q$, the $Q$ independent signatures and the nonzero projected calibration derivative give $n_\nu=2Q+2$, so $\operatorname{rank}\mathbf{J}_{\mathrm{e}}\le2Q-2$ by Theorem~\ref{thm:rank}(b). \hfill$\blacksquare$

\section{Proof of Proposition~\ref{prop:coop} and Corollary~\ref{cor:singleport}}\label{app:coop}

Let $c_\alpha=2|\alpha|^2/\sigma^2$ and $\mathcal{E}=\sum_{\ell\in\mathcal{L}_+}\mathcal{E}_\ell$, and let the sums over $\boldsymbol{\chi}_\ell$ run over $\mathcal{L}_+$. For $Q=1$, \eqref{eq:efim_proj} reads $\mathbf{J}_{\mathrm{e}}=c_\alpha\operatorname{Re}\{\mathbf{D}^H\mathbf{D}-\mathbf{D}^H\mathbf{s}\mathbf{s}^H\mathbf{D}/\|\mathbf{s}\|^2\}$. By orthogonality, $\mathbf{D}^H\mathbf{D}=\sum_\ell\mathbf{D}_\ell^H\mathbf{D}_\ell$, $\|\mathbf{s}\|^2=\mathcal{E}$, and $\mathbf{D}^H\mathbf{s}=\sum_\ell\mathbf{D}_\ell^H\mathbf{s}_\ell=\mathcal{E}\bar{\boldsymbol{\chi}}$, while the definition of $\mathbf{P}^\perp_{\mathbf{s}_\ell}$ gives $c_\alpha\operatorname{Re}\{\mathbf{D}_\ell^H\mathbf{D}_\ell\}=\mathbf{J}^{\mathrm{loc}}_\ell+c_\alpha\,\mathcal{E}_\ell\operatorname{Re}\{\boldsymbol{\chi}_\ell\boldsymbol{\chi}_\ell^H\}$ for $\ell\in\mathcal{L}_+$ and $c_\alpha\operatorname{Re}\{\mathbf{D}_\ell^H\mathbf{D}_\ell\}=\mathbf{J}^{\mathrm{loc}}_\ell$ otherwise. Hence
\begin{equation}
\mathbf{J}_{\mathrm{e}}=\sum_\ell\mathbf{J}^{\mathrm{loc}}_\ell+c_\alpha\operatorname{Re}\Big\{\sum_\ell\mathcal{E}_\ell\boldsymbol{\chi}_\ell\boldsymbol{\chi}_\ell^H-\mathcal{E}\bar{\boldsymbol{\chi}}\bar{\boldsymbol{\chi}}^H\Big\},
\end{equation}
and since $\sum_\ell\mathcal{E}_\ell\boldsymbol{\chi}_\ell=\mathcal{E}\bar{\boldsymbol{\chi}}$, the matrix in braces equals $\sum_\ell\mathcal{E}_\ell(\boldsymbol{\chi}_\ell-\bar{\boldsymbol{\chi}})(\boldsymbol{\chi}_\ell-\bar{\boldsymbol{\chi}})^H$, which proves \eqref{eq:coop}--\eqref{eq:jcoop}. Each summand of \eqref{eq:jcoop} is the real part of a positive semidefinite Hermitian matrix, and $\sum_\ell\mathbf{J}^{\mathrm{nuis}}_\ell-\mathbf{J}^{\mathrm{coop}}=c_\alpha\,\mathcal{E}\operatorname{Re}\{\bar{\boldsymbol{\chi}}\bar{\boldsymbol{\chi}}^H\}\succeq\mathbf{0}$, which gives the two-sided bound. For real $\mathbf{v}$, $\mathbf{v}^\top\operatorname{Re}\{\mathbf{x}\mathbf{x}^H\}\mathbf{v}=|\mathbf{v}^\top\mathbf{x}|^2$, so the quadratic form vanishes if and only if $\mathbf{v}^\top\boldsymbol{\chi}_\ell=\mathbf{v}^\top\bar{\boldsymbol{\chi}}$ for every $\ell\in\mathcal{L}_+$.

For Corollary~\ref{cor:singleport}, with $\mathbf{R}_x$ diagonal, the rows of $\mathbf{X}$ are mutually orthogonal, so the components of $\mathbf{y}$ generated by distinct feed ports are orthogonal, and those observed at distinct output ports occupy disjoint entries. Link $\ell=(f,o)$ contributes $\mathbf{s}_\ell=a_\ell\mathbf{r}_\ell$ with $\mathbf{r}_\ell=[\mathbf{X}]_{f,:}^\top\otimes\mathbf{e}_o$ independent of $\mathbf{p}$ and $\|\mathbf{r}_\ell\|^2=T[\mathbf{R}_x]_{ff}$, so $\mathbf{D}_\ell=\mathbf{r}_\ell(\nabla_{\mathbf{p}}a_\ell)^\top$. Since $a_\ell\neq0$, $\mathbf{P}^\perp_{\mathbf{s}_\ell}\mathbf{r}_\ell=\mathbf{0}$ gives $\mathbf{J}^{\mathrm{loc}}_\ell=\mathbf{0}$, $\mathcal{E}_\ell=T[\mathbf{R}_x]_{ff}|a_\ell|^2$, and $\boldsymbol{\chi}_\ell=(\nabla_{\mathbf{p}}a_\ell)^*/a_\ell^*=(\nabla_{\mathbf{p}}\ln a_\ell)^*=\boldsymbol{\tau}_\ell-j\boldsymbol{\phi}_\ell$. Since $\operatorname{Re}\{(\mathbf{x}-j\mathbf{y})(\mathbf{x}-j\mathbf{y})^H\}=\mathbf{x}\mathbf{x}^\top+\mathbf{y}\mathbf{y}^\top$ for real $\mathbf{x},\mathbf{y}$, \eqref{eq:jcoop} becomes \eqref{eq:scatter}. \hfill$\blacksquare$

\section{Proof of Proposition~\ref{prop:loss}}\label{app:loss}

For the tail bound, by \eqref{eq:order}, $z_n\ge z_{N+1}+(n-N-1)d_{\min}$ for $n\ge N+1$. With \eqref{eq:b} and $\rho_n=\kappa(1-\kappa)^{n-1}$,
\begin{align}
|h^{(N')}-h^{(N)}|&\le\bar g\sum_{n=N+1}^{N'}\sqrt{\kappa}(1-\kappa)^{\frac{n-1}{2}}e^{-\alpha_gz_n/2}\notag\\
&\le\sqrt{\kappa}\,\bar g\,(1-\kappa)^{\frac N2}e^{-\frac{\alpha_gz_{N+1}}{2}}\notag\\
&\;\times\sum_{j\ge0}\big(\sqrt{1-\kappa}\,e^{-\alpha_gd_{\min}/2}\big)^j,
\end{align}
which is \eqref{eq:tail}. The same argument bounds the tail of the position derivatives. Under M-III, differentiating a guided coefficient with respect to $\alpha_g$ or $\beta_g$ multiplies it by $-z_n/2$ or $-jz_n$, which are bounded by the guide length, so the calibration derivatives obey the same tail bound. By the product rule in \eqref{eq:y}, the full real Jacobians at prefix lengths $N$ and $N'$ therefore differ by at most a constant times $(1-\kappa)^{N/2}e^{-\alpha_g z_{\min,N+1}/2}$. The lower bound $g_0$ bounds the nuisance pseudoinverses uniformly, so the range projectors differ by at most $2/g_0$ times the nuisance Jacobian difference, and \eqref{eq:efim_factor} gives the same geometric order for the EFIM difference. The identity $\mathbf{A}^{-1}-\mathbf{B}^{-1}=\mathbf{A}^{-1}(\mathbf{B}-\mathbf{A})\mathbf{B}^{-1}$ with $\lambda_0$ transfers it to the inverse EFIMs, whose target-block traces are bounded away from zero because the Jacobians are bounded, so taking square roots and the maximum over targets preserves the bound.

For the attenuation sensitivity, with $\partial b_{f,n}/\partial\alpha_g=-\tfrac12z_{f,n}b_{f,n}$, $\partial[\mathbf{h}^T]_f/\partial\alpha_g=-\tfrac12\bar z^T_f[\mathbf{h}^T]_f$, and likewise at the receiver, so $\partial\ln a_\ell/\partial\alpha_g=-\tfrac12\zeta_\ell$. Hence $\partial\mathcal{E}_\ell/\partial\alpha_g=2\mathcal{E}_\ell\operatorname{Re}\{\partial\ln a_\ell/\partial\alpha_g\}=-\mathcal{E}_\ell\operatorname{Re}\zeta_\ell$ and $\partial\boldsymbol{\chi}_\ell/\partial\alpha_g=(\nabla_{\mathbf{p}}\partial\ln a_\ell/\partial\alpha_g)^*=-\tfrac12(\nabla_{\mathbf{p}}\zeta_\ell)^*$. Differentiate $\mathbf{M}=\sum_\ell\mathcal{E}_\ell\tilde{\boldsymbol{\chi}}_\ell\tilde{\boldsymbol{\chi}}_\ell^H$ with $\tilde{\boldsymbol{\chi}}_\ell=\boldsymbol{\chi}_\ell-\bar{\boldsymbol{\chi}}$. The terms containing $\partial\bar{\boldsymbol{\chi}}$ are multiplied by $\sum_\ell\mathcal{E}_\ell\tilde{\boldsymbol{\chi}}_\ell=\mathbf{0}$ and vanish, so
\begin{align}
\partial\mathbf{M}=\sum_\ell\partial\mathcal{E}_\ell\,\tilde{\boldsymbol{\chi}}_\ell\tilde{\boldsymbol{\chi}}_\ell^H
+\sum_\ell\mathcal{E}_\ell\big[\partial\boldsymbol{\chi}_\ell\,\tilde{\boldsymbol{\chi}}_\ell^H+\tilde{\boldsymbol{\chi}}_\ell\,\partial\boldsymbol{\chi}_\ell^H\big],
\end{align}
Substituting the two derivatives, writing $\operatorname{Re}\zeta_\ell=\bar\zeta+(\operatorname{Re}\zeta_\ell-\bar\zeta)$ in the first sum, and using $\mathbf{J}_{\mathrm{e}}=c_\alpha\operatorname{Re}\mathbf{M}$ from Corollary~\ref{cor:singleport} gives \eqref{eq:dJdalpha}. Its first term $-\bar\zeta\mathbf{J}_{\mathrm{e}}$ contributes $\bar\zeta\,\mathrm{PEB}^2$ to $\partial\mathrm{PEB}^2/\partial\alpha_g$. \hfill$\blacksquare$

\section{Proofs of Lemma~\ref{lem:rank1} and Proposition~\ref{prop:alg}}\label{app:rank1}

SINR feasibility with $\gamma_k>0$ and $\sigma_k^2>0$ forces $\mathbf{h}_k^\top\mathbf{Q}_k\mathbf{h}_k^*>0$, so \eqref{eq:rank1_rec} is well defined and $\tilde{\mathbf{Q}}_k$ has rank one. The Cauchy--Schwarz inequality in the semi-inner product $\langle\mathbf{x},\mathbf{y}\rangle=\mathbf{x}^H\mathbf{Q}_k\mathbf{y}$ gives $|\mathbf{x}^H\mathbf{Q}_k\mathbf{h}_k^*|^2\le(\mathbf{x}^H\mathbf{Q}_k\mathbf{x})(\mathbf{h}_k^\top\mathbf{Q}_k\mathbf{h}_k^*)$, i.e., $\mathbf{x}^H(\mathbf{Q}_k-\tilde{\mathbf{Q}}_k)\mathbf{x}\ge0$ for any $\mathbf{x}$, hence $\tilde{\mathbf{S}}\succeq\mathbf{S}\succeq\mathbf{0}$. Since $\sum_k\tilde{\mathbf{Q}}_k+\tilde{\mathbf{S}}=\mathbf{R}_x$, the FIM (Lemma~\ref{lem:efim}), the objective, and the feed powers are unchanged. In \eqref{eq:sinr}, the useful power $\mathbf{h}_k^\top\tilde{\mathbf{Q}}_k\mathbf{h}_k^*=\mathbf{h}_k^\top\mathbf{Q}_k\mathbf{h}_k^*$ and the interference-plus-noise $\mathbf{h}_k^\top(\mathbf{R}_x-\tilde{\mathbf{Q}}_k)\mathbf{h}_k^*+\sigma_k^2$ are also unchanged, and so is every SINR. The optimal value of the relaxation lower-bounds that of (P1-in) and is attained by a point satisfying the rank constraints, which proves Lemma~\ref{lem:rank1}.

For Proposition~\ref{prop:alg}, $\mathbf{z}^{(0)}$ is feasible with $f<\infty$, every candidate lies in the feasible interval of its coordinate given the others, and a candidate is accepted only if its value is below $(1-\epsilon_{\mathrm{a}})f_{\mathrm{cur}}<\infty$. Hence every iterate satisfies \eqref{eq:order} with finite $f$, and $f_{\mathrm{cur}}$ changes only by decreasing. The returned inner variables solve (P1-in) at the final $\mathbf{z}$ and satisfy the rank constraints by Lemma~\ref{lem:rank1}. To see that $f_{\inf}>0$, note that all elements are at height $H>0$, so every $\|\mathbf{M}_i\|$ is bounded over $\mathcal{Z}$, and $\operatorname{tr}\mathbf{R}_x\le\sum_mP_m$. By \eqref{eq:fim_linear}, $\mathbf{J}_{\mathrm{e}}\preceq\mathbf{J}_{pp}\preceq c_{\max}\mathbf{I}$ uniformly, so every finite value satisfies $\operatorname{tr}([\mathbf{J}_{\mathrm{e}}^{-1}]_q)\ge2/c_{\max}$. After $a$ acceptances, $f_{\inf}\le f_{\mathrm{cur}}<(1-\epsilon_{\mathrm{a}})^af(\mathbf{z}^{(0)})$, which bounds $a$. Every sweep either contains an acceptance or halves $\Delta$, which can happen at most $\lceil\log_2(\Delta_0/\Delta_{\min})\rceil+1$ times before the loop exits, and a sweep evaluates at most six candidates per coordinate. Finally, the loop exits only after a sweep without acceptance, in which every candidate of every coordinate was evaluated at the then-current $\mathbf{z}$, and the inner variables are optimal by convexity of the relaxation and Lemma~\ref{lem:rank1}. \hfill$\blacksquare$


\bibliographystyle{IEEEtran}
\bibliography{refs}

@article{fukuda2022pinching,
  author  = {A. Fukuda and H. Yamamoto and H. Okazaki and Y. Suzuki and K. Kawai},
  title   = {Pinching antenna: Using a dielectric waveguide as an antenna},
  journal = {NTT DOCOMO Tech. J.},
  volume  = {23}, number = {3}, pages = {5--12}, month = jan, year = {2022}
}

@article{ding2025flexible,
  author  = {Z. Ding and R. Schober and H. V. Poor},
  title   = {Flexible-antenna systems: A pinching-antenna perspective},
  journal = {IEEE Trans. Commun.},
  volume  = {73}, number = {10}, pages = {9236--9253}, month = oct, year = {2025},
  doi     = {10.1109/TCOMM.2025.3555866}
}

@article{liu2026survey,
  author  = {Y. Liu and H. Jiang and X. Gan and X. Xu and J. Guo and Z. Wang and C. Ouyang and X. Mu and Z. Ding and A. Nallanathan and O. A. Dobre and G. K. Karagiannidis and R. Schober},
  title   = {A survey of pinching-antenna systems ({PASS})},
  journal = {IEEE Commun. Surveys Tuts.},
  volume  = {28}, pages = {7250--7280}, year = {2026},
  doi     = {10.1109/COMST.2026.3731727}
}

@article{xu2026attenuation,
  author  = {Y. Xu and Z. Ding and R. Schober and T.-H. Chang},
  title   = {Pinching-antenna systems with in-waveguide attenuation: Performance analysis and algorithm design},
  journal = {IEEE Trans. Wireless Commun.},
  volume  = {25}, pages = {14564--14580}, year = {2026},
  doi     = {10.1109/TWC.2026.3676036}
}

@article{ouyang2026swan,
  author  = {C. Ouyang and H. Jiang and Z. Wang and Y. Liu and Z. Ding},
  title   = {Uplink and downlink communications in segmented waveguide-enabled pinching-antenna systems ({SWANs})},
  journal = {IEEE Trans. Commun.},
  volume  = {74}, pages = {3688--3703}, year = {2026},
  doi     = {10.1109/TCOMM.2026.3653885}
}

@article{wang2025sensing,
  author  = {Z. Wang and C. Ouyang and Y. Liu and A. Nallanathan},
  title   = {Wireless sensing via pinching-antenna systems},
  journal = {IEEE Wireless Commun. Lett.},
  volume  = {14}, number = {11}, pages = {3475--3479}, month = nov, year = {2025},
  doi     = {10.1109/LWC.2025.3595160}
}

@article{ding2025crlb,
  author  = {Z. Ding},
  title   = {Pinching-antenna assisted {ISAC}: A {CRLB} perspective},
  journal = {npj Wireless Technology},
  volume  = {1}, number = {1}, pages = {Art. no. 4}, month = nov, year = {2025},
  doi     = {10.1038/s44459-025-00004-z}
}

@article{mao2026multi,
  author  = {W. Mao and Y. Lu and Y. Xu and B. Ai and O. A. Dobre and D. Niyato},
  title   = {Multi-waveguide pinching antennas for {ISAC}},
  journal = {IEEE Trans. Wireless Commun.},
  volume  = {25}, pages = {5846--5858}, year = {2026},
  doi     = {10.1109/TWC.2025.3621316}
}

@article{jiang2026bcrb,
  author  = {H. Jiang and C. Ouyang and Z. Wang and Y. Liu and A. Nallanathan and Z. Ding},
  title   = {Pinching-antenna-assisted sensing: A {B}ayesian {C}ram{\'e}r--{R}ao bound perspective},
  journal = {IEEE Trans. Commun.},
  volume  = {74}, pages = {9075--9092}, year = {2026},
  doi     = {10.1109/TCOMM.2026.3690382}
}

@article{li2026pass_isac,
  author  = {H. Li and R. Zhong and Z. Pan and C. Dong and J. Lei and Y. Liu},
  title   = {Pinching antenna systems for integrated sensing and communications},
  journal = {IEEE Trans. Wireless Commun.},
  volume  = {25}, pages = {13416--13429}, year = {2026},
  doi     = {10.1109/TWC.2026.3668822}
}

@misc{geng2026swan,
  author       = {Y. Geng and T. H. Cheng and K. C. Teh and Z. Ding},
  title        = {{CRLB} minimization for {ISAC} systems with segmented waveguide-enabled pinching antenna},
  howpublished = {arXiv:2604.00572}, year = {2026}
}

@misc{bozanis2026secure,
  author       = {D. Bozanis and V. K. Papanikolaou and S. A. Tegos and P. D. Diamantoulakis and C. Masouros and G. K. Karagiannidis},
  title        = {Secure near-field {ISAC} with pinching-antenna systems: Exploiting constructive-destructive interference},
  howpublished = {arXiv:2609.32715}, year = {2026}
}

@misc{pakravan2026dual,
  author       = {S. Pakravan and I. Trigui and W. Ajib and W.-P. Zhu},
  title        = {Joint pinching-antenna position and power optimization for dual-waveguide {ISAC} systems},
  howpublished = {arXiv:2610.11987}, year = {2026}
}

@misc{yang2026joint,
  author       = {H. Yang and H. Feng and E. Bedeer and M. Zeng and M. Wang and G. Huang and M. Huang},
  title        = {Joint antenna geometry and transmit covariance design for near-field multicast {ISAC} with pinching antenna arrays},
  howpublished = {arXiv:2609.23693}, year = {2026}
}

@article{gan2026cpass,
  author  = {X. Gan and Y. Liu},
  title   = {Center-fed pinching antenna system ({C-PASS}): Modeling, analysis, and beamforming design},
  journal = {IEEE Trans. Wireless Commun.},
  volume  = {25}, pages = {21911--21925}, year = {2026},
  doi     = {10.1109/TWC.2026.3718454}
}

@article{liu2022isac,
  author  = {F. Liu and Y. Cui and C. Masouros and J. Xu and T. X. Han and Y. C. Eldar and S. Buzzi},
  title   = {Integrated sensing and communications: Toward dual-functional wireless networks for {6G} and beyond},
  journal = {IEEE J. Sel. Areas Commun.},
  volume  = {40}, number = {6}, pages = {1728--1767}, month = jun, year = {2022},
  doi     = {10.1109/JSAC.2022.3156632}
}

@article{behdad2024multistatic,
  author  = {Z. Behdad and {\"O}. T. Demir and K. W. Sung and E. Bj{\"o}rnson and C. Cavdar},
  title   = {Multi-static target detection and power allocation for integrated sensing and communication in cell-free massive {MIMO}},
  journal = {IEEE Trans. Wireless Commun.},
  volume  = {23}, number = {9}, pages = {11580--11596}, month = sep, year = {2024},
  doi     = {10.1109/TWC.2024.3383209}
}

@article{wang2026cfisac,
  author  = {Z. Wang and V. W. S. Wong and R. Schober},
  title   = {Cooperative {ISAC} for joint localization and velocity estimation in cell-free {MIMO} systems},
  journal = {IEEE J. Sel. Areas Commun.},
  volume  = {44}, pages = {642--658}, year = {2026},
  doi     = {10.1109/JSAC.2025.3613283}
}

@article{zhu2024movable,
  author  = {L. Zhu and W. Ma and R. Zhang},
  title   = {Movable antennas for wireless communication: Opportunities and challenges},
  journal = {IEEE Commun. Mag.},
  volume  = {62}, number = {6}, pages = {114--120}, month = jun, year = {2024},
  doi     = {10.1109/MCOM.001.2300212}
}

@article{wong2021fluid,
  author  = {K.-K. Wong and A. Shojaeifard and K.-F. Tong and Y. Zhang},
  title   = {Fluid antenna systems},
  journal = {IEEE Trans. Wireless Commun.},
  volume  = {20}, number = {3}, pages = {1950--1962}, month = mar, year = {2021},
  doi     = {10.1109/TWC.2020.3037595}
}

@article{ma2024ma_sensing,
  author  = {W. Ma and L. Zhu and R. Zhang},
  title   = {Movable antenna enhanced wireless sensing via antenna position optimization},
  journal = {IEEE Trans. Wireless Commun.},
  volume  = {23}, number = {11}, pages = {16575--16589}, month = nov, year = {2024},
  doi     = {10.1109/TWC.2024.3443293}
}

@article{shen2010fundamental,
  author  = {Y. Shen and M. Z. Win},
  title   = {Fundamental limits of wideband localization--{P}art {I}: A general framework},
  journal = {IEEE Trans. Inf. Theory},
  volume  = {56}, number = {10}, pages = {4956--4980}, month = oct, year = {2010},
  doi     = {10.1109/TIT.2010.2060110}
}

@article{godrich2010target,
  author  = {H. Godrich and A. M. Haimovich and R. S. Blum},
  title   = {Target localization accuracy gain in {MIMO} radar-based systems},
  journal = {IEEE Trans. Inf. Theory},
  volume  = {56}, number = {6}, pages = {2783--2803}, month = jun, year = {2010},
  doi     = {10.1109/TIT.2010.2046246}
}

@article{liu2022crb,
  author  = {F. Liu and Y.-F. Liu and A. Li and C. Masouros and Y. C. Eldar},
  title   = {{C}ram{\'e}r-{R}ao bound optimization for joint radar-communication beamforming},
  journal = {IEEE Trans. Signal Process.},
  volume  = {70}, pages = {240--253}, year = {2022},
  doi     = {10.1109/TSP.2021.3135692}
}

@article{mityagin2015zero,
  author  = {B. S. Mityagin},
  title   = {The zero set of a real analytic function},
  journal = {Math. Notes},
  volume  = {107}, number = {3--4}, pages = {529--530}, month = mar, year = {2020},
  doi     = {10.1134/S0001434620030189}
}

@misc{he2026crb,
  title         = {{Cram\'er-Rao} Bound Analysis for Cell-Free {ISAC} Systems with Fluid Intelligent Metasurfaces},
  author        = {Changhao He and Jiguang He and Asmaa Abdallah and Ahmed M. Eltawil},
  year          = {2026},
  eprint        = {2608.16634},
  archivePrefix = {arXiv},
  primaryClass  = {cs.IT},
  howpublished  = {arXiv:2608.16634},
  url           = {https://arxiv.org/abs/2608.16634}
}

\end{document}